%% file: master.tex
\documentclass[12pt]{article}
\usepackage{times}
\usepackage{epsfig}
\usepackage{aaspp4}
\usepackage{astrobib}

\def \lsim {\lesssim}
\def \gsim {\gtrsim}
\def \figwidth {0.9 \linewidth}

\begin{document}
\input aas_journals

\input href_linked
\psrotatefirst

\title{A Photometric and Weak Lensing Analysis of the 
$z=0.42$ Supercluster MS0302+17}
\author{
N.~Kaiser\altaffilmark{1}, 
G.~Wilson\altaffilmark{1}, 
G.~Luppino\altaffilmark{1}, 
L.~Kofman\altaffilmark{1}, 
I.~Gioia\altaffilmark{1},
M.~Metzger\altaffilmark{2}
and
H.~Dahle\altaffilmark{1}
}
\altaffiltext{1}{Institute for Astronomy, University of Hawaii}
\altaffiltext{2}{Caltech}
\authoraddr{Institute for Astronomy, University of Hawaii, 2680 Woodlawn Drive, Honolulu, Hawaii 96822, 
email: kaiser@hawaii.edu}

\input abstract

\clearpage
\tableofcontents
\listoffigures
\listoftables
\clearpage

\input introduction

\input xrayprops

\input photometry

\input lensing

\input discussion

\input acknowledge

\bibliographystyle{apj}
\bibliography{astro,clusters,weaklensing,kaiser_ref,kaiser_nonref}


\end{document}

%% file: aas_journals.tex
%
%
%



\def\aj{{AJ}}                   
\def\araa{{ARA\&A}}             
\def\apj{{ApJ}}                 
\def\apjl{{ApJ}}                
\def\apjs{{ApJS}}               
\def\ao{{Appl.~Opt.}}           
\def\apss{{Ap\&SS}}             
\def\aap{{A\&A}}                
\def\aapr{{A\&A~Rev.}}          
\def\aaps{{A\&AS}}              
\def\azh{{AZh}}                 
\def\baas{{BAAS}}               
\def\jrasc{{JRASC}}             
\def\memras{{MmRAS}}            
\def\mnras{{MNRAS}}             
\def\pra{{Phys.~Rev.~A}}        
\def\prb{{Phys.~Rev.~B}}        
\def\prc{{Phys.~Rev.~C}}        
\def\prd{{Phys.~Rev.~D}}        
\def\pre{{Phys.~Rev.~E}}        
\def\prl{{Phys.~Rev.~Lett.}}    
\def\pasp{{PASP}}               
\def\pasj{{PASJ}}               
\def\qjras{{QJRAS}}             
\def\skytel{{S\&T}}             
\def\solphys{{Sol.~Phys.}}      
\def\sovast{{Soviet~Ast.}}      
\def\ssr{{Space~Sci.~Rev.}}     
\def\zap{{ZAp}}                 
\def\nat{{Nature}}              
\def\iaucirc{{IAU~Circ.}}       
\def\aplett{{Astrophys.~Lett.}} 
\def\apspr{{Astrophys.~Space~Phys.~Res.}}
\def\bain{{Bull.~Astron.~Inst.~Netherlands}} 
\def\fcp{{Fund.~Cosmic~Phys.}}  
\def\gca{{Geochim.~Cosmochim.~Acta}}   
\def\grl{{Geophys.~Res.~Lett.}} 
\def\jcp{{J.~Chem.~Phys.}}      
\def\jgr{{J.~Geophys.~Res.}}    
\def\jqsrt{{J.~Quant.~Spec.~Radiat.~Transf.}}
\def\memsai{{Mem.~Soc.~Astron.~Italiana}}
\def\nphysa{{Nucl.~Phys.~A}}   
\def\physrep{{Phys.~Rep.}}   
\def\physscr{{Phys.~Scr}}   
\def\planss{{Planet.~Space~Sci.}}   
\def\procspie{{Proc.~SPIE}}   

\let\astap=\aap
\let\apjlett=\apjl
\let\apjsupp=\apjs
\let\applopt=\ao

%% file: href_linked.tex

\newcommand{\astrophref}[1]{\href{http://xxx.lanl.gov/abs/astro-ph/#1}{\tt{http://xxx.lanl.gov/abs/astro-ph/#1}}}
\newcommand{\myhref}[1]{\href{#1}{\tt{#1}}}

%% file: abstract.tex
\begin{abstract}{ 
We perform a weak lensing and photometric study of the $z=0.42$ supercluster 
MS0302+17 using deep $I$ and $V$ band images taken with the UH8K CCD mosaic 
camera at the CFHT. We use archival ROSAT HRI data to estimate fluxes, gas 
masses and, in one case, the binding mass of the three major clusters.
We then use our CCD data to determine the optical richness and luminosities
of the clusters and to map out the spatial distribution of the early type
galaxies in the supercluster and in other foreground and background
structures.
We measure the gravitational shear from a sample of $\simeq 30,000$ faint
background galaxies in the range $22 \lsim m_I \lsim 26$ and find this 
correlates strongly with that predicted from the early type galaxies if they 
trace the mass with $M/L_B \simeq 250 h$.
We make 2-dimensional reconstructions of the
mass surface density.  These recover all of the major concentrations
of galaxies and indicate that most of the supercluster mass, like the 
early type galaxies, is concentrated in the three X-ray clusters, and we obtain
mean mass-to-light ratios for the clusters of $M/L_B \simeq 260 h$.
Cross-correlation of the measured mass surface density
with that predicted from the early
type galaxy distribution shows a strong peak at zero lag (significant at
the $\simeq 9$-sigma level), and that at separations $\gsim 200 h^{-1}$kpc
the early galaxies trace the mass very accurately.  This conclusion is 
supported by cross-correlation in Fourier space; we see little evidence
for any variation of $M/L$ or `bias' with scale, and from the longest
wavelength modes with $\lambda = 1.5-6 h^{-1}$Mpc we find 
$M/L \simeq (280 \pm 40)h$, quite similar to that obtained for the cluster 
centers.
We discuss the implication of these results for the cosmological
density parameter.
}\end{abstract}

\keywords{Cosmology: Observations, Cosmology: Dark Matter, Cosmology:
Gravitational Lensing, Galaxies: Clusters, Galaxies: Photometry}

%% file: introduction.tex
\section{Introduction}
\label{sec:introduction}

Weak lensing, in the sense of the statistical distortion of the shapes of 
faint background galaxies, has now been measured for
quite a number of clusters:
\citeNP{twv90};
\citeNP{bfkms93};
\citeNP{fksw94};
\citeNP{bmf94};
\citeNP{dml94};
\citeNP{mdfb94};
\citeNP{fm94};
\citeNP{sd95};
\citeNP{tf95};
\citeNP{fmdbk96};
\citeNP{skss96};
\citeNP{bs97};
\citeNP{sedcosb97};
\citeNP{fbrt97};
\citeNP{ft97},
and provides a powerful and direct measurement of the
total mass distribution in clusters.
Following the pioneering attempt of \citeN{tvjm84}
using photographic plates,  several groups 
(\citeNP{bbs96}; \citeNP{dt96}; \citeNP{gcir96}; \citeNP{hsdk98})
have reported measurements of the `galaxy-galaxy lensing' effect
due to dark halos around galaxies,
and there have also been attempts to estimate the shear due to large-scale structure
(\citeNP{vjt83},
\citeNP{mbvbssk94},
\citeNP{svmjsf98}).
Most of these studies have been made with fairly small
CCD detectors; this severely limits the distance out to which one
can probe the cluster mass distribution and also limits the precision
of galaxy-galaxy lensing and large-scale shear studies.

Here we present a weak lensing analysis of deep $I$ and $V$ photometry of the 
field containing the $z \simeq 0.42$ supercluster
MS0302+17 taken with a large 8192$\times$8192 pixel CCD mosaic camera
(the UH8K; \citeNP{lmkcgm95}) mounted behind the prime focus
wide field corrector on the CFHT.  
The field of view of this camera on this telescope measures
$0^\circ.5$ on a side, and greatly increases the range of accessible scales, 
and should yield improved
precision for large-scale and galaxy-galaxy shear measurements.
Our goal here is to explore the total (dark plus luminous) 
mass distribution in this supercluster field and see how this relates to the
distribution of super-cluster galaxies and the X-ray emitting gas.

The target field, centered roughly on 
$\rm{RA} = 3^h 5^m 24^s.0, \rm{DEC} = 17^\circ 18' 0''.0$, (J2000)
contains three prominent clusters in
a supercluster at $z \simeq 0.42$.
The first of these clusters to be found, CL0303+1706, 
was detected optically by \citeN{dg92}. It has a redshift of 
$z = 0.418$, a velocity dispersion $\sigma_v = 912$ km/s, and
was the the target of an Einstein IPC pointed observation. 
This observation revealed two sources within $\sim 1'$ of the position given by
Dressler and Gunn, one of which, it is now apparent, coincides 
with densest and most massive concentration of galaxies in the cluster, and
which has an X-ray luminosity $L_{\rm X} \simeq 3.0 h_{50}^{-2}\times 10^{44}$ erg/s.
Analysis of the IPC image and optical follow up
by the EMSS team \cite{emss90,emss91,gl94} revealed the
presence of the two neighboring clusters
MS0302+1659 and MS0302+1717 with redshifts $z= 0.426, 0.425$  and
X-ray luminosities 
$L_{\rm X} \simeq 5.0 \times 10^{44}h_{50}^{-2}$erg/s $4.3\times 10^{44}h_{50}^{-2}$ erg/s.
The fluxes here are taken from the analysis of the IPC images by \citeN{fbm94},
who also obtained redshifts for a large number of galaxies in the the two 
X-ray brighter clusters,
and obtained velocity dispersions of and 921 and 821 km/s respectively.  
\citeN{cnoc96}  have reported a velocity dispersion of 646 km/s 
for MS0302+1659 based on
27 redshifts and \citeN{cnoc97} have presented an extended catalog of
redshifts for this cluster.
Deep optical CCD images of  MS0302+1659 were taken by
\citeN{mfmps92} which revealed the presence of a pair of 
giant arcs and additional photometry was taken by
\citeN{giraud92} who found evidence for variability of features in 
one of the arcs.

The angular separations
of the clusters CL0303+1706, MS0302+1659 and MS0302+1717 (which we
will refer to as CL-E, CL-S, CL-N respectively)
are 
$\theta_{\rm NE}= 17'.64$, $\theta_{\rm NS}= 18'.49$, and $\theta_{\rm SE}=14'.04$.  
Assuming an Einstein de Sitter universe, the angular diameter distance is
$D = 2 c H_0^{-1} (1 - (1+z)^{-1/2}) / (1 + z)$ so $D(z=0.42) \simeq 679 h^{-1}$Mpc and 
the physical scale is 
3.29$h^{-1}$ kpc/arcsec or 197$h^{-1}$ kpc/arcmin.  For pure Hubble flow the line of sight physical
separation is $\Delta r^\parallel = (1 + z)^{-5/2} c \Delta z / H_0$ or about $1.24 \Delta z h^{-1}$Gpc.  
Fabricant {\it et al.\/} found $\Delta z_{\rm NS} = 0.0009 \pm 0.0012$ from which we infer that
the separation of the EMSS clusters in redshift space is $\Delta r_{\rm NS} \simeq 3.8 h^{-1} {\rm Mpc}$, and is
primarily transverse to the line of sight 
($\Delta r^\parallel_{\rm NS} \simeq 1.1\pm1.5 h^{-1} {\rm Mpc}$, $\Delta r^\perp_{\rm NS} \simeq 3.6 h^{-1} {\rm Mpc}$).
The Dressler-Gunn cluster E lies about $\Delta r^\parallel_{\rm NE} \simeq 9 h^{-1} {\rm Mpc}$ in front
of the two EMSS clusters. 
The parallel distance components may be somewhat 
distorted by peculiar motions.  For an empty open cosmology 
$D = 0.5 c H_0^{-1} (1 - (1+z)^{-2})$ so $D(z=0.42) \simeq 756 h^{-1}$Mpc and 
$\Delta r^\parallel = (1 + z)^{-2} \Delta z c / H_0$ or about $1.49 \Delta z h^{-1}$Gpc, in which case
the transverse and line of sight physical dimensions are larger than the EdS values by about
$11$\% and $20$\% respectively.
For an extreme $\Lambda$ dominated model with $\Omega_\Lambda = 1$, $\Omega_m = 
0$,
$D = c z H_0^{-1} / (1 + z)$ and so $D(z=0.42) \simeq 887 h^{-1}$Mpc
and $\Delta r^\parallel = (1 + z)^{-1} \Delta z c / H_0$ or about $2.11 \Delta z
 h^{-1}$Gpc, in which case
the transverse and line of sight physical dimensions are larger than the EdS 
values by about
$30$\% and $69$\% respectively.
The parallel distance components may be somewhat
distorted by peculiar motions.  

This field was chosen for weak lensing as the
high velocity dispersions and high
X-ray luminosities of these clusters (along with the presence of arcs in
one case)
lead one to suspect that they might be potent lenses.
Also, the whole system
fits conveniently within
the approximately $30' \times 30'$ field of view of the
UH8K camera.  This system provides an example of a quasi-linear
system in the process of formation and it is of interest to see if
weak lensing can detect, or constrain, the presence of `bridges' of
dark matter that can be expected between clusters \cite{bkp96}.  However,
our main goal is to explore the correlation between mass and 
optical luminosity
density to obtain estimates of the mass to luminosity ratio $M/L$,
or equivalently the inverse of the `bias factor',
(using the methodology of \citeN{k91}, \citeN{schneider98}) and to see if
this varies with scale. A source of systematic uncertainty in weak lensing mass
estimates is the still largely unknown redshift distribution of the
faint background galaxies. As a check we compare
our mass estimates for the centers of the clusters with mass estimates from
the velocity dispersions and, in one case, an estimate of the X-ray gas temperature.

The outline of the paper is as follows: 
In \S\ref{sec:xrayprops} we review the existing X-ray and 
optical spectroscopy results.  We obtain improved fluxes and 
positions from archival ROSAT HRI observations. We make crude estimates of the
central gas masses and, for CL-S estimate the central total
mass from an ASCA temperature obtained by \citeN{henry98} and compare with the
optical velocity dispersion.
In \S\ref{sec:photometry} we explore the optical structure of the supercluster. 
We briefly describe the image processing; present a color image of the field;
and show the size-magnitude distribution for the $\sim 44,000$ objects detectable
at $> 6$-sigma significance. We describe the somewhat non-standard photometric system 
we are using, and compute expected colors as a function of redshift and galaxy type. 
We plot the position, color and luminosities of the brighter galaxies, and 
we compute counts of galaxies as a function of magnitudes for 
regions around the clusters and for the complementary field region.
In the absence of extensive
redshift information it is hard to
accurately measure the total optical luminosity distribution due to the
confusing effect of foreground structures.  If we restrict attention to the
narrow band of $V-I$ color occupied by the early type galaxies at the
supercluster redshift this foreground clutter is effectively removed and
we obtain a cleaner picture of the distribution of these galaxies and
obtain estimates of the optical richness and
luminosity functions for the three X-ray bright clusters.

In \S\ref{sec:shearanalysis} we present the weak lensing analysis.
In \S\ref{sec:galaxyselection} we describe how we select the
faint galaxies used to measure the shear and how
we calibrate the relation between the observed image polarization and
the gravitational shear.  The calibration and also the correction
for point spread function anisotropy is described more fully 
elsewhere \cite{k98}.
Our shear analysis features an optimized weighting scheme, and we show how the
weight is distributed in magnitude and on the size-magnitude plane.
In \S\ref{sec:kappapred} we predict the  dimensionless surface density
$\kappa(\theta) = \Sigma_m / \Sigma_{\rm crit}$ assuming that
early type galaxy light traces the mass with some a constant $M/L$.  We do this
both for a narrow slice around the supercluster redshift and also for
a broader range of $V-I$ color, corresponding to a range of redshifts $0.2 \lsim z \lsim 0.6$,
and which includes a significant contribution from a foreground structure at $z \simeq 0.3$.
In \S\ref{sec:shearanal} we correlate the measured shear with that predicted from the
early type galaxy distribution.  
In \S\ref{sec:reconstruction} we
perform 2-dimensional mass reconstructions and compare these with
the constant $M/L$ prediction. In \S\ref{sec:clusterml} we estimate 
central mass-to-light ratios for the three X-ray bright clusters, and
we compare these with velocity dispersions and X-ray temperature information.
In \S\ref{sec:crosscorrelation} we compute the real space light-mass 
cross-correlation function to obtain a more precise estimate of the
$M/L$ parameter and we compare the profile of the mass-light cross-correlation
function and that of the luminosity auto-correlation function.
We also perform the cross-correlation in Fourier
space in order to see if there is any evidence for `scale dependent
biasing' (i.e. variation of the $M/L$ parameter with scale).

%% file: xrayprops.tex
\section{X-Ray Properties}
\label{sec:xrayprops}

In this section we review the published X-ray 
results and  obtain improved fluxes and 
positions from archival ROSAT HRI observations. 
We make crude estimates of the
central gas masses and, for CL-S estimate the central total
mass from an ASCA temperature obtained by Henry and compare with the
optical velocity dispersion.

\begin{figure}[htbp!]
\centering\epsfig{file=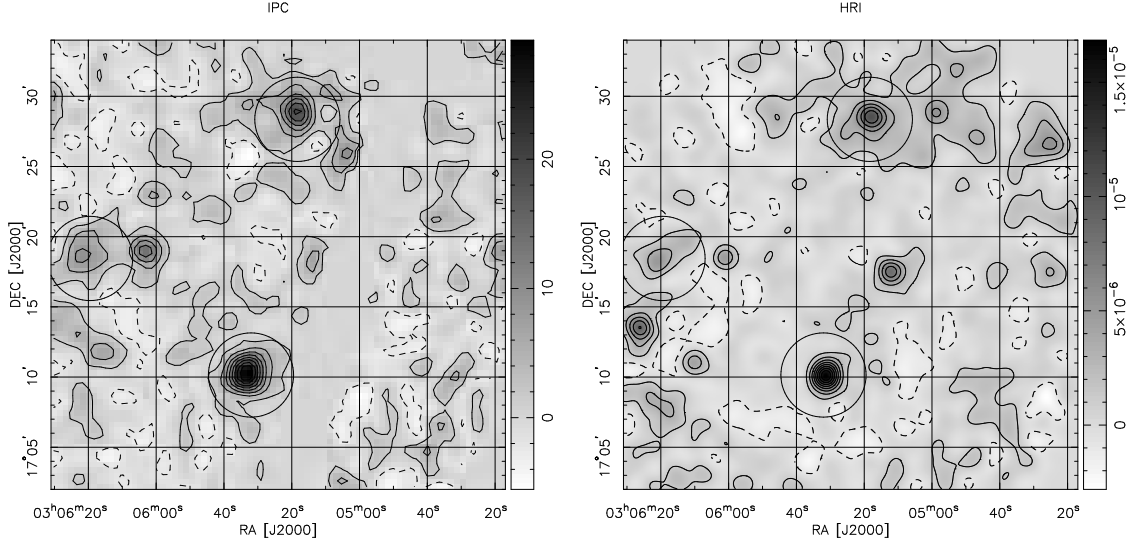,width=\figwidth,angle=-90}
\caption[IPC and HRI images.]{left hand panel
shows the IPC X-ray image from the Einstein
archive.  The units of intensity
are in IPC counts per second per square arcmin. The circles
show the optical cluster centroids. Right hand panel shows a
composite of the two HRI pointings, smoothed with a $35''$ Gaussian.
The intensity units are HRI counts per second per $(5'')^2$
pixel.
}
\label{fig:xray}
\end{figure}

The Einstein IPC image from the Einstein database is
shown in the left hand panel of figure \ref{fig:xray},  
with the circles indicating the locations of the clusters.
Only the region where we have optical photometry coverage is shown.
Analysis of these
data by \citeN{fbm94} gave estimates of the luminosities
for these clusters quoted in the Introduction.  
The identification of the target cluster was originally somewhat
confused as the cluster is quite extended, and
the location provided by Dressler and Gunn
was displaced somewhat to the West from what is now seen to be the
cluster center, and lay right between the two sources on the left
with $\delta \simeq 17^\circ 19'$.  There is some indication of
extended emission around CL-N, but the situation is somewhat confused
by the shadowing of the support structure.

\begin{figure}[htbp!]
\centering\epsfig{file=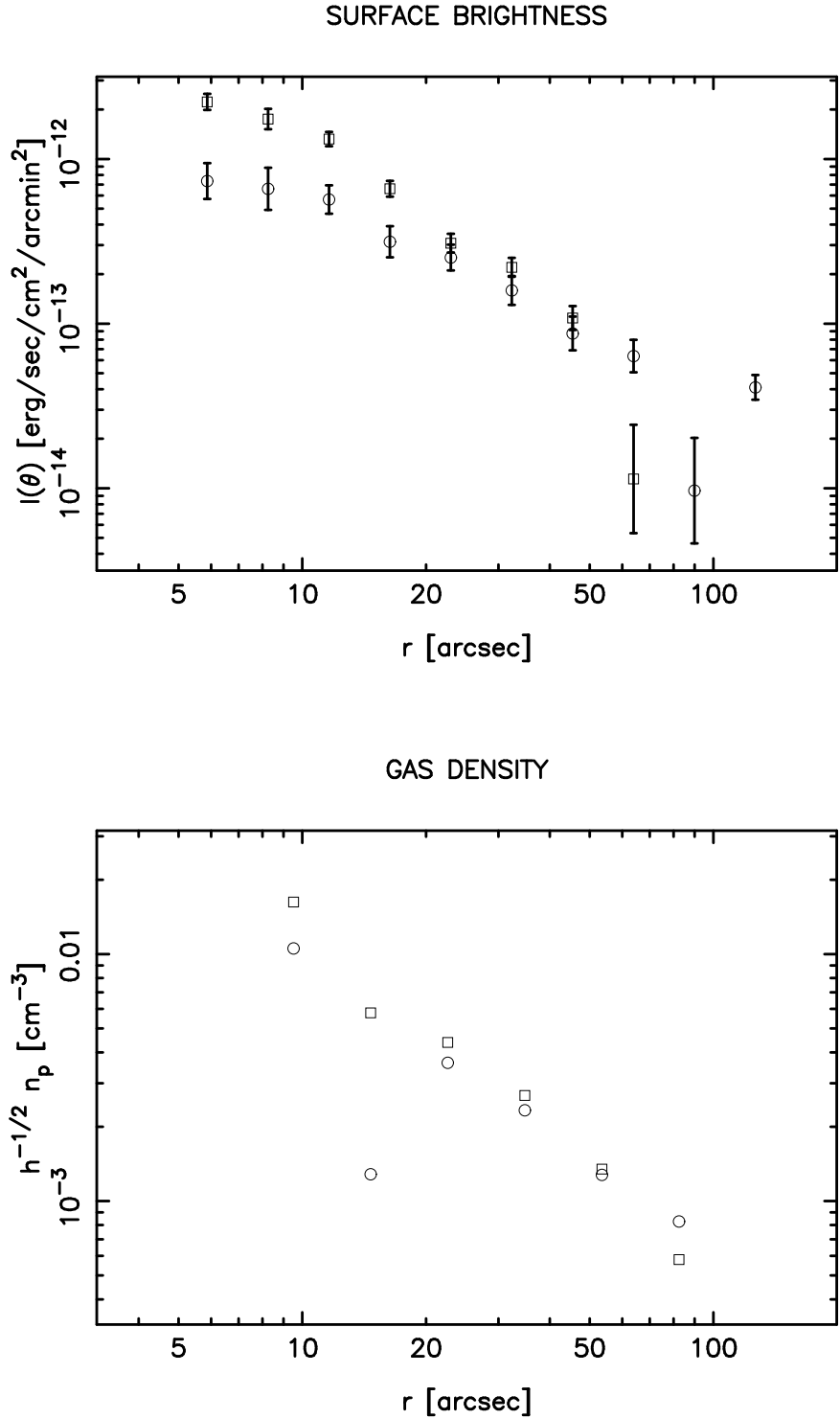,width={0.7 \linewidth},angle=0}
\caption[X-ray luminosity and gas density profiles]{Upper panel
shows the azimuthally averaged surface brightness profiles
for the clusters CL-S (squares) and CL-N (circles). CL-E is not shown
as it is clearly non-circular.  
Lower panel shows the
gas density profile obtained by direct deprojection as described in
the text.  
}
\label{fig:xrayprofile}
\end{figure}

There have been two further observations of the field with
the ROSAT HRI which are now in the public domain.    
One of these was of $\sim 3.5 \times 10^4$s integration time and was
centered on CL-S, but also contains CL-E,
while the other of $\sim 9 \times 10^4$s integration
contains all three clusters.  These have higher resolution and
higher astrometric precision than the IPC data.
To analyze these data we used Steve Snowden's
software \cite{snowden98} to generate a model for the particle background 
for each pointing, which we
subtract, and also to generate a detailed exposure map
which we divide into the subtracted image to correct for
vignetting.  We then mask out objects detected at $> 4 $-sigma
significance level and take the mean of what is left to obtain
an estimate of the sky background, which we also subtract.
The neutral hydrogen column density along the line of
sight is $N_{H} = 1\times 10^{21}$cm$^{-2}$ \cite{dl90} 
from which we find (assuming a Raymond-Smith spectrum with observed 
temperature of $3.43$keV and $0.25$ times solar metallicity) 
that one HRI count per second corresponds to
an unabsorbed bolometric flux of $1.06\times 10^{-10}$ erg/sec/cm$^2$
(for the often quoted passbands of $0.2-4.5$keV and $0.5-2.0$keV
the corresponding conversion factors are 
$7.75\times 10^{-11}$ erg/sec/cm$^2$ and
$3.79\times 10^{-11}$ erg/sec/cm$^2$).
A smoothed surface brightness image is shown in
figure \ref{fig:xrayprofile}.
The  unsmoothed
HRI image shows that most of the other sources are still unresolved
(and are therefore most probably not emission from cluster gas).
The smoothed HRI image also shows the extended emission around CL-N, though now
without the masking effect of the IPC ribs. 
CL-E is seen to be extended towards the NW, 
and at higher resolution appears to have a second core.  This bimodality
is also seem in the highly non-circular distribution of cluster galaxies
so it would appear that this is most probably a merger event.
The azimuthally averaged surface brightness profiles for CL-N, CL-S are
shown in figure \ref{fig:xrayprofile}.  
CL-N appears to have a slope with index $\gamma \sim -1$ (where
$I(\theta) \propto \theta^\gamma$) with significant emission extending to
$\theta \sim 100-200''$.  CL-S has a slightly steeper slope $\gamma
\sim -1.4$ within $\theta \simeq 50''$, outside of which radius the
emission drops sharply and there is no statistically significant flux 
detectable at larger
radii. This cluster is remarkably compact as compared to more typical bright
X-ray clusters which have cores with essentially flat emission profiles
extending to about this radius and with most of the gas residing at larger
radii.

Assuming spherical symmetry, the
rest-frame surface brightness at impact parameter $r$ is
\begin{equation}
I_{\rm rest}(r) = {1 \over 4 \pi} \int dz \; 
\epsilon_{\rm ff}(\sqrt{r^2 + z^2})
\end{equation}
where the Bremsstrahlung emissivity is
$\epsilon_{\rm ff} = \alpha_{\rm ff} n_p^2 T^{1/2}$  where
$\alpha_{\rm ff} \simeq 3\times 10^{-27}$erg cm$^3$ s$^{-1}$ K$^{-1/2}$
\cite{sarazin88}.
Approximating the integral as a sum over cubical `voxels',
the observed surface brightness for a pixel at
radius $i$ pixels from the cluster center in an image with pixel
size  $\Delta\theta$
is 
\begin{equation}
I_{\rm obs}(i) = {a_0 \omega_l \Delta \theta \over 4 \pi (1 + z)^5} 
\sum\limits_j F(\sqrt{i^2 + j^2})
\end{equation}
with $F(i) \equiv \epsilon_{\rm ff}(a_l \omega_l i \Delta \theta)$.
We can invert the projection sum numerically 
to obtain $F(i) = 4 \pi (1 + z)^5 (a_0 \omega_l \Delta \theta)^{-1} 
\tilde I(i)$
where $I(i) \equiv \sum \tilde I(\sqrt{i^2 + j^2})$, and hence the gas density:
\begin{equation}
n_p(r) = \left({4 \pi (1 + z)^5 \tilde I(r/a_l \omega_l \Delta \theta) 
\over a_0 \omega_l \Delta \theta T^{1/2} \alpha_{\rm ff}}\right)^{1/2}
\end{equation}
The gas density profiles are shown in the lower panel of figure 
\ref{fig:xrayprofile} from which we estimate
gas masses within radius corresponding to $1'.5$ of
$M_S \simeq 2.5 \times 10^{12} h^{-5/2} M_\odot$,
$M_N \simeq 2.8 \times 10^{12} h^{-5/2} M_\odot$.

\citeN{henry98} has obtained
an ASCA spectrum for CL-S from which he finds a rest-frame temperature
of $4.6 \pm 0.8$keV. Assuming isothermality,
the total mass interior to $r$ is $M(r) = - (k T r / G \mu m_p) d \ln n / d \ln r$,
and using a logarithmic slope of $d \ln n / d \ln r = -1.5$ and a
mean molecular weight $\mu = 0.6$
we find $M(<r = 167h^{-1} {\rm kpc}) 
= 4.25\times 10^{13} h^{-1} M_\odot$, this
radius corresponding to the angular scale of $50''$.
Combining these results, we find a gas fraction for CL-S of
$M_{\rm gas} / M_{\rm tot} \simeq 0.06 h^{-3/2}$.  Both the gas and total
mass estimates should be considered to be lower bounds, as there may
be emission at larger radii which we cannot detect and one could
also imagine that the cluster has an extended dark matter halo
which these central X-ray observations cannot constrain.
It is interesting to compare $(kT / G \mu m_p)^{1/2} \simeq 
854$km/s with the observed velocity dispersion of
$921$km/s.  As we shall see, the galaxies in the
cluster are, like the observable gas, highly concentrated and have a very similar scale
length, so
it is reassuring that the kinetic energy per unit mass is
essentially the same for the gas and for the galaxies.  The high temperature
and very small size of the cluster indicate a very short
sound crossing time $t = r / c_s \simeq 2 \times 10^8$yr.

The inferred gas density in the very center of CL-S 
(at around $15''$ or about $50 h^{-1}$kpc say) is $n \simeq
1.4\times 10^{-2} h^{-1/2}{\rm cm}^{-3}$, implying a cooling
rate of about $\Gamma \sim 1.4 \times 10^{-18} h^{-1/2} {\rm s}^{-1}$ or
a cooling time $t_{\rm cool} \sim 2.2 \times 10^9 h^{1/2} {\rm yr}$,
as compared to the age of the Universe at $z = 0.42$ of
$t \simeq 3.9 \times10^9 h^{-1} {\rm yr}$ for $\Omega = 1$, or
$t \simeq 7.1 \times10^9 h^{-1} {\rm yr}$ for an open $\Omega \simeq 0$
cosmology, so the ratio of the cooling time to the age of the
Universe $t_{\rm cool} / t(z = 0.42)$ is in the range  $(0.3-0.5)h^{3/2}$
so we are dealing with a strong, but still quasi-static, cooling flow.  We estimate the gas loss rate
as $\dot M \sim 100-200 M_\odot$/yr.
At a radius of $50''$ the
density is an order of magnitude lower, 
so the cooling time is an order
of magnitude larger and, for $h = 0.7$,  $t_{\rm cool} / t(z = 0.42)$
lies in the range 1.7-3.0 so the cooling is somewhat larger than the age of the
Universe. However, for low $\Omega$, and in the absence of heat sources,  much of the
gas we see will by now have dropped out of the cooling flow.

The HRI derived positions, count rates, fluxes luminosities
of the three clusters are summarized in
table \ref{tab:xrayprops}.  The agreement between our luminosities and
those of \citeN{fbm94} is generally
very good.  For CL-S, for instance, and for the $0.2-4.5$keV band,
we find $L = 4.9\pm0.4 \times h_{50}^{-2} 10^{44}$erg/s whereas they find
$L = 5.0\pm0.8 \times h_{50}^{-2} 10^{44}$ erg/s.  

\input table_xrayprops

%% file: table_xrayprops.tex
\suppressfloats[t]
\begin{table*}[htbp!]
\begin{center}
\begin{tabular}{cccccc}
\hline\hline
Cluster	& RA(J2000) & DEC(J2000) & $r$ [cts/s] & $f_{\rm bol}$ [erg/s/cm$^2$] & $h_{50}^2 L_{\rm bol}$ [erg/s] \\
\hline
CL-S & $3^h05^m31^s.49$ & $17^\circ 10'16''.32$	& $7.1\pm0.5\times 10^{-3}$ & $7.6\pm0.6\time 10^{-13}$ & $6.8\pm0.5\times 10^{44}$	\\
CL-E & $3^h06^m18^s.98$ & $17^\circ 18'33''.91$	& $3.8\pm0.7\times 10^{-3}$ & $4.0\pm0.8\time 10^{-13}$ & $3.6\pm0.7\times 10^{44}$ \\
CL-N & $3^h05^m17^s.81$ & $17^\circ 28'37''.64$	& $5.6\pm0.6\times 10^{-3}$ & $6.0\pm0.7\time 10^{-13}$ & $5.4\pm0.6\times 10^{44}$	\\
\hline
\end{tabular}
\caption[Cluster X-ray properties]{Cluster X-ray Properties. Positions are locations of peaks of the smoothed HRI
image. Count rates, fluxes and luminosities are 
for $1'.5$ radius apertures centered on the
given position, except for CL-E where we include the
flux in a second aperture centered on the secondary feature to the NW of the 
main cluster.}
\label{tab:xrayprops}
\end{center}
\end{table*}

%% file: photometry.tex
\section{Optical Photometry}
\label{sec:photometry}

\subsection{Data Acquisition and Processing.}

In September 1995, we obtained deep images 
of the MS0302 supercluster using the UH8K CCD camera mounted at the 
prime focus of the CFHT.
At an image scale of $\simeq 0''.207$/pixel, the 8192$\times$8192 pixel CCD mosaic
spans a field of view of $\sim$29$'$ on a side. 
The images were taken in superb observing conditions: photometric
sky with  $\sim 0''.6$ seeing.  
Multiple 15 minute exposures were taken, with the telescope shifted
slightly ($\sim 20-30''$) between exposures, in order to facilitate
flat fielding and the construction of a seamless final image. In all,
11 I-band images and 6 V-band images were used to give final
exposure times of the summed images of 9900s and 5400s respectively.

The image processing was rather involved, and is described
more fully elsewhere \cite{kwld98}  but essentially consists of
four phases; pre-processing of original images; registration to
obtain the transformation from chip to celestial coordinates; 
warping and averaging to produce the final images
(and also auxiliary files describing the variation of photon counting
noise, point spread function shape etc); and finally the generation
of catalogs of objects.

In the pre-processing we first generated a median sky image 
or `super-flat' for each chip
and each passband and divided this into the data images.  The super-flat 
was also used to identify bad columns, traps, and
other defects, which were flagged as unreliable.  To further suppress
residual variations in the sky background level
we then subtracted from the images a smooth
local sky background estimate formed from the depths of the minima of 
a slightly smoothed version of the
super-flat divided data images.

In the registration phase we took the locations of stars measured on the
data chips, along with stars with known accurate celestial coordinates
from the USNOA catalog \cite{usnoa} and solved in a least square manner for a set
of low order polynomials --- one for each of the $2{\rm K}\times 4{\rm K}$ images ---
which describe the mapping from rectilinear chip coordinates to
celestial coordinates.  We computed a sequence of progressive 
refinements to the transformation.
The final solution gave transformations with a precision of 
about 5 milli-arcsec, which
accurately correct for field distortion, atmospheric refraction, 
and the layout of the chips in the detector, and also
any other distortions produced by inhomogeneities of the filters and/or
mechanical distortion of the detector.

In the image mapping phase we applied the spatial transformations to generate
a quilt of slightly overlapping images covering a planar projection of the 
celestial sphere. We chose the `orthographic', shape-preserving
projection \cite{gc95} with tangent point at $(\alpha, \delta)_{\rm J2000} =
(3^h05^m24^s.0, 17^\circ 18'0''.0)$  and rotation angle of $2^\circ.635$
(this being chosen to render the star trails aligned with the vertical axis
of the final image to facilitate masking of these artifacts). 
Cosmic ray pixels and satellite trails etc.~were identified
by comparing each image section with the median of a stack of images
taken in the same passband, and were flagged as unreliable. 
The final images were made by simple averaging of the non-flagged pixels
in the stack of warped images.
The combination of
the mosaic chip geometry layout, the pattern of shifts for the exposures, and
the presence of unreliable data of various types resulted in a somewhat
non-uniform sky noise level, which we kept track of by making an auxiliary
image of the sky noise.  
The FWHM of stars in these stacked images were approximately $0''.60$.
We also generated a detailed model for how the
point spread function (psf) shape varies across the final image (this is
quite complicated as the psf shape changes discontinuously
across chip boundaries) and we apply this below to correct the measured
galaxy shapes to what they would have been if measured with a telescope with
perfectly circular psf. 

\begin{figure}[htbp!]
\centering\epsfig{file=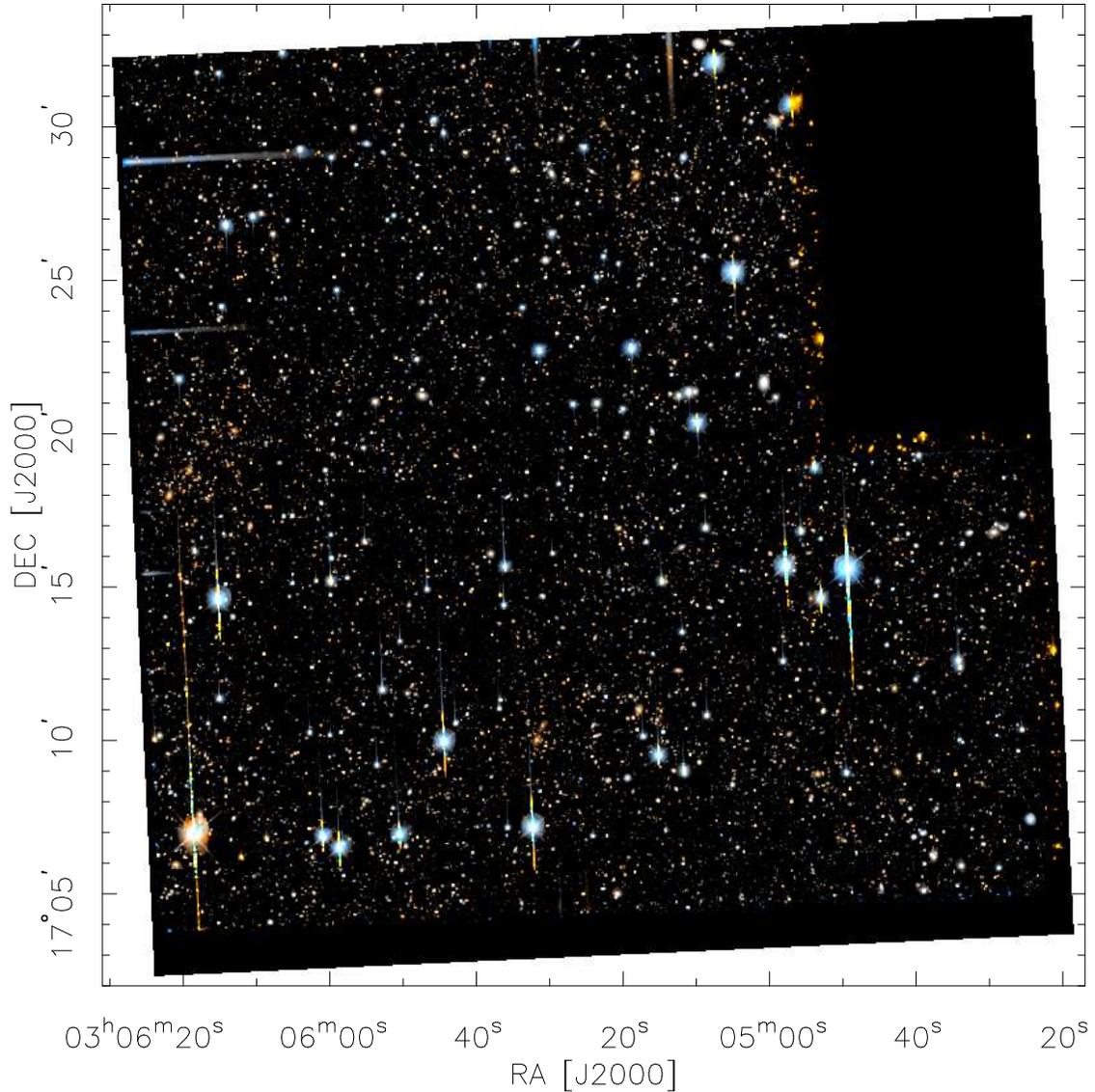,width=\figwidth}
\caption[Color image of the supercluster field.]{Color image generated from the summed V and I band images.
This is a heavily reduced (de-magnified) image of the field derived from the
final averaged images (which have a re-sampled pixel scale of $0''.15$ and cover
$12,000^2$ pixels).  In order to show both the faint background galaxies and
the much brighter foreground galaxies,
the brightness of each pixel here was taken to be proportional to the logarithm of the sum of the
two V and I-band images, and the individual intensities are such that the red and blue signals
are in the same proportion as in the linear I, V images (with a constant multiplicative scaling
applied to the blue value) and the green signal is the mean of the
red and blue signals. Thus this is not a true 3-color image (which would
look much whiter), but is useful to distinguish the different
galaxy types present. }
\label{fig:colorimage}
\end{figure}

A color image generated from the summed V and I band exposures is shown in
figure \ref{fig:colorimage}.
The Dressler-Gunn cluster CL-E is clearly visible
close to the Eastern edge of the field
as an agglomeration of early-type galaxies with very similar color.
The two X-ray selected clusters can be seen fairly clearly, and the giant
arcs in the Southern cluster are apparent.  Both of these clusters contain
galaxies with the same color as in CL-E, but in both cases there are
also bright bluer galaxies in the vicinity, 
suggesting that there is some foreground contamination.

\begin{figure}[htbp!]
\centering\epsfig{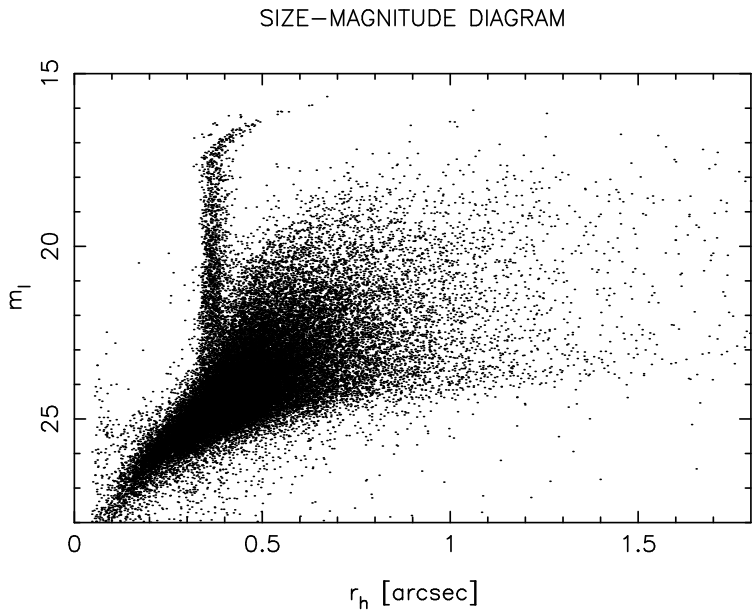}
\caption[Size-magnitude diagram.]{Size-magnitude diagram for the 
combined sample. The vertical stellar locus is clearly seen, and stars can clearly
be separated from galaxies for $m_I \lsim 23$. At fainter magnitudes the stars 
and galaxies merge, but the stars are then a small contamination.
}
\label{fig:sizemag}
\end{figure}

After registering and summing the images we ran our `hierarchical
peak finding algorithm' 
to detect objects (\citeNP{ksb95}, \citeNP{kwld98}) on each of the I- and V-band 
images and then ran our aperture
photometry and shapes analysis routines to compute magnitudes, radii
and other properties of the objects.  For each of the $I,V$ catalogs we also
estimated magnitudes using the summed image for the other passband
(but with the
same center and aperture radius) in order to provide 
accurate color information.
We also constructed a combined `I+V' catalog in which, if an
object was detected in both V and I then we include only the most
significant detection, whereas if the object was only detected in one
passband then the information for that filter only is included. This
is the primary catalog we use in the analysis below.
To remove spurious detections from the diffraction spikes 
around bright stars and trails caused by 
poor charge transfer for saturated pixels, we generated a mask consisting of 
the rectangles shown in figure \ref{fig:faintgals}, and objects lying
within these rectangles were not used in the following analysis.
The combined catalog contains about 44,000 objects detected at
$> 6$-sigma significance.  The total unmasked area is 
$A = 0.165$ square degrees, so the number density of these objects is
about 74 per square arcmin. The size-magnitude
distribution for the combined catalog is shown in figure \ref{fig:sizemag},
from which it can be seen that
the $6$-sigma significance threshold of the catalog
corresponds to $m_I\simeq 25.5$ for
point-like objects --- though with incompleteness setting in at brighter
magnitudes for extended objects. 

\subsection{Photometric System and Galactic Extinction}

The photometric system used here, whose passbands
we denote by $B', V', R', I'$, and which differ
somewhat from the standard $B,V,R,I$ passbands, is described in detail 
elsewhere \cite{kwld98}. There we compute the redshift and
SED dependent $k$-terms in the transformation from UH8K apparent
magnitude $m_{F'}$ to standard passband absolute magnitude $M_F$:
\begin{equation}
M_F - c_F = m_{F'} - c_F + k_{FF'}(z; {\rm type}) - 
5 \log_{10} (D_l / 10{\rm pc}) 
\end{equation}  
Here $c_F$ is the magnitude of Vega in passband $F$.
We also show that one can form
synthetic magnitudes as a linear combination of our UH8K $I', V'$
magnitudes 
$m_{F_z} = a_{F_z} (m_{I'} - c_{I}) + (1 - a_{F_z}) (m_{V'} - c_{V})$
for $F = B,V\ldots$, whose transformation to 
standard rest-frame absolute magnitudes are
\begin{equation}
M_F = m_{F_z} + k_{F_z}(z) - 5 \log_{10} (D_l / 10{\rm pc}) - c_F
\end{equation}
and where the coefficients $a_{F_z}$ have been chosen so that the
$k$ term is essentially independent of spectral type for galaxies
at the redshift of the clusters $z = 0.42$. 
For an Einstein - de Sitter Universe,
and with $h = 1$,
the distance modulus is $5 \log_{10} (D_l / 10{\rm pc}) = 40.68$.

The apparent magnitudes here have been corrected for galactic
extinction using the \citeN{sfd98} IRAS-based prediction.  
Their maps give
$E(B-V) = 0.125, 0.134, 0.113$ for the clusters N,E,S respectively. 
These are similar, and we adopt an average value of $E(B-V) = 0.124$,
from which we obtain corrections of
$\Delta m_I = 0.243, \;  \Delta m_V = 0.402$. 

\begin{figure}[htbp!]
\centering\epsfig{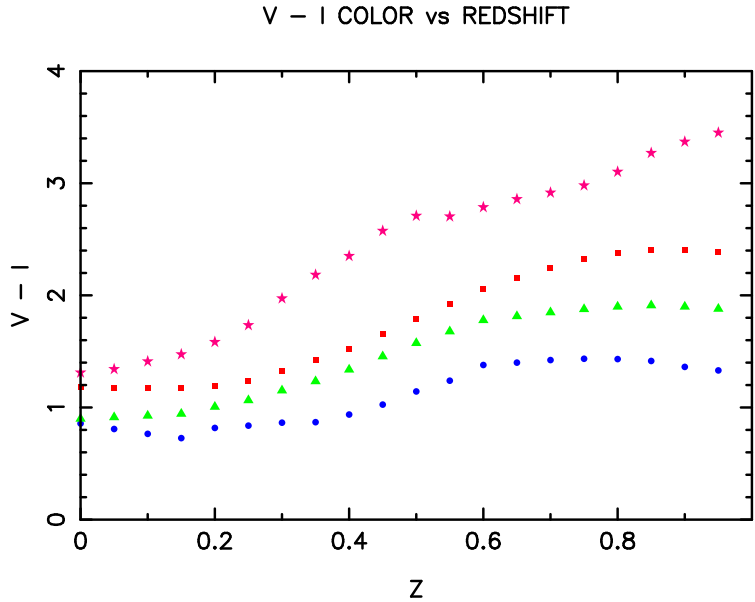}
\caption[Galaxy colors vs redshift.]{Galaxy color vs redshift for the 
UH8K system and using broad band SED's from 
\citeN{cww80} . Symbols encode the galaxy type as follows:
E0 = stars; Scd = squares; Sbc = triangles; Im = circles.
}
\label{fig:galcolorvsz}
\end{figure}

In the lensing analysis we shall use the distribution of bright galaxies
to predict the projected dimensionless surface density and thereby the
gravitational shear.  In doing this it is important to
discriminate between structures at
low and high redshift (as the lensing signal derives
largely from the latter).  A useful discriminant, especially for early
type galaxies, is the $V'-I'$ color, which is plotted as a function of
redshift for various galaxy types in figure \ref{fig:galcolorvsz}.

\subsection{Bright Galaxy Properties and Cluster Luminosities}

We will now explore the properties of the brighter galaxies which define
the supercluster and
estimate the optical luminosities of the clusters.  In the absence of
redshift information
this is rather difficult because of the presence of foreground
structures.  We can however obtain fairly accurate estimates of the
individual cluster luminosities for the early-type cluster
galaxies as the noise due to foreground and background contamination
is then greatly reduced.

\begin{figure}[htbp!]
\centering\epsfig{file=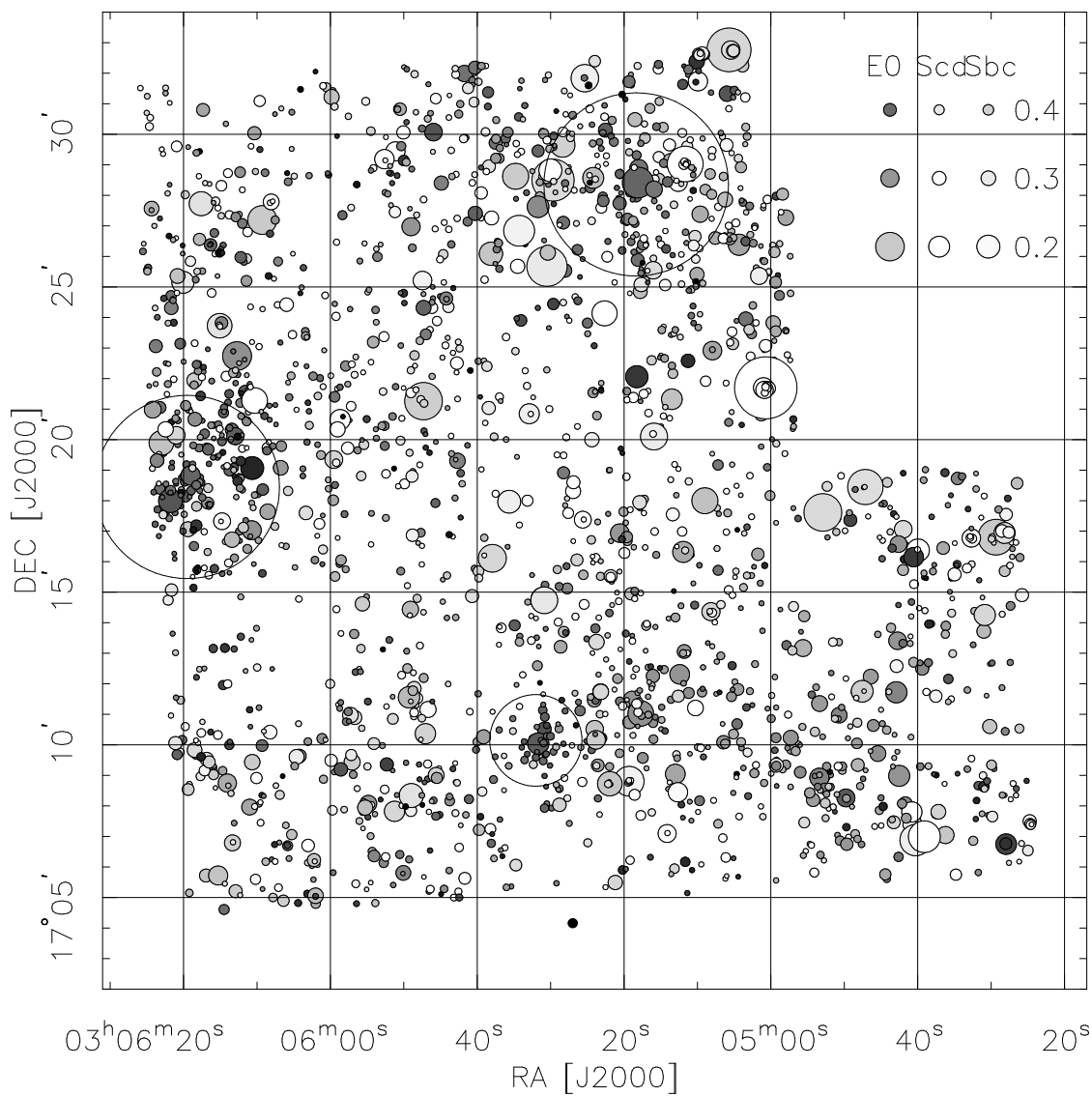,width=\figwidth,angle=-90}
\caption[Spatial distribution of bright galaxies.]{Plot of all 
galaxies brighter than $I=21$ where the area of each
circle is proportional to the flux density of the galaxy, and
the shade of the disc encodes the $V-I$ color, ranging from
$V-I = 1$ for a white disc to $V-I = 3$ for a black disc. 
The legend in the upper right corner shows the image that would
be produced by an $L_*$ galaxy ($M_B = -19.68$) of various
types at redshifts $0.2, 0.3, 0.4$.
Stars have been removed.  The three main galaxy concentrations are
apparent, as is an extension to the West from  
southern EMSS cluster.  The large circles have radius $3'$ ($1'.5$ for CL-S) and are
centered on the X-ray cluster positions.
}
\label{fig:brightgals}
\end{figure}

In figure \ref{fig:brightgals} we plot the
locations, colors, and fluxes of the bright galaxies in the
field. 
At the center of each circle drawn around the 
X-ray positions resides a dense concentration of very red galaxies
with colors as expected for elliptical galaxies at $z = 0.42$.
In addition to the red elliptical galaxies, a large number of
bluer and brighter galaxies can be seen.
One conspicuous concentration lies $2'-3'$ to the East of CL-N, and, to
judge from the colors and fluxes of the galaxies, lies at $z \simeq 0.2$.
Another, more diffuse concentration, with colors and luminosities
indicative of early type galaxies at $z \simeq 0.3$ can be seen extending
Westward  of CL-S. We have obtained redshifts for two of the bright galaxies
in this complex and find that they do indeed lie at $z \simeq 0.3$.

\begin{figure}[htbp!]
\centering\epsfig{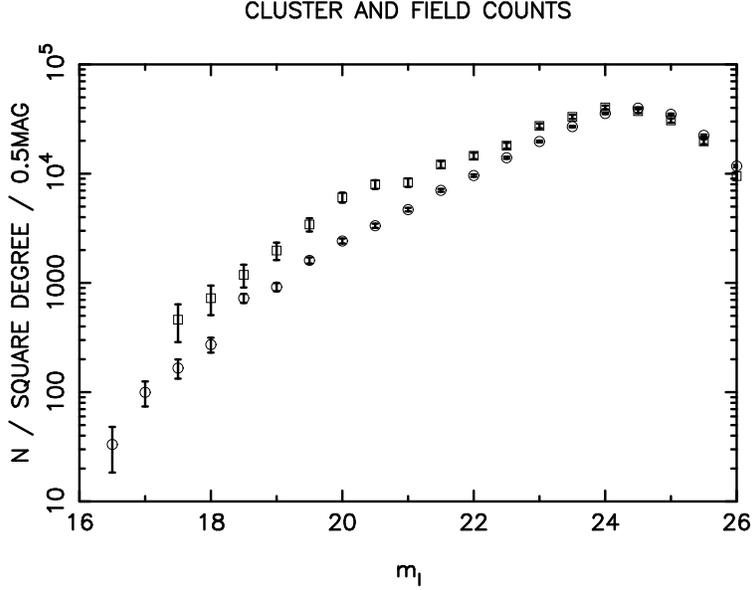}
\caption[Galaxy counts]{Counts of galaxies for the `cluster' regions defined by the
circles in figure \ref{fig:brightgals}, and for the complementary
region. 
}
\label{fig:counts}
\end{figure}

We can obtain an unbiased, if somewhat noisy,  
estimate of the abundance of galaxies in general
in the three X-ray clusters by computing the counts of all galaxies
(the number of galaxies per solid angle per magnitude interval) within the
cluster-centered circles and then subtracting the `background'
counts obtained for the complementary region.  The counts as
a function of $I$-band apparent magnitude are shown
in figure \ref{fig:counts}.  In computing the counts we allow for the
somewhat irregular survey geometry caused by our masking of bright 
stars (the mask is shown below in figure \ref{fig:faintgals}).  
To estimate the unmasked areas we count the number of objects from a 
densely sampled random Poissonian distribution 
of points which was masked in exactly the same way as the galaxy
catalogs.  The error bars in figure \ref{fig:faintgals} are
poissonian only, and the real fluctuations are somewhat larger.
The excess of galaxy counts due to the clusters is clearly visible.
Similar results are obtained for the synthetic rest-frame $B$-magnitude
$m_{B_z} \simeq (m_I + m_V) / 2$, though these start to become incomplete
at somewhat brighter magnitudes. 

\begin{figure}[htbp!]
\centering\epsfig{file=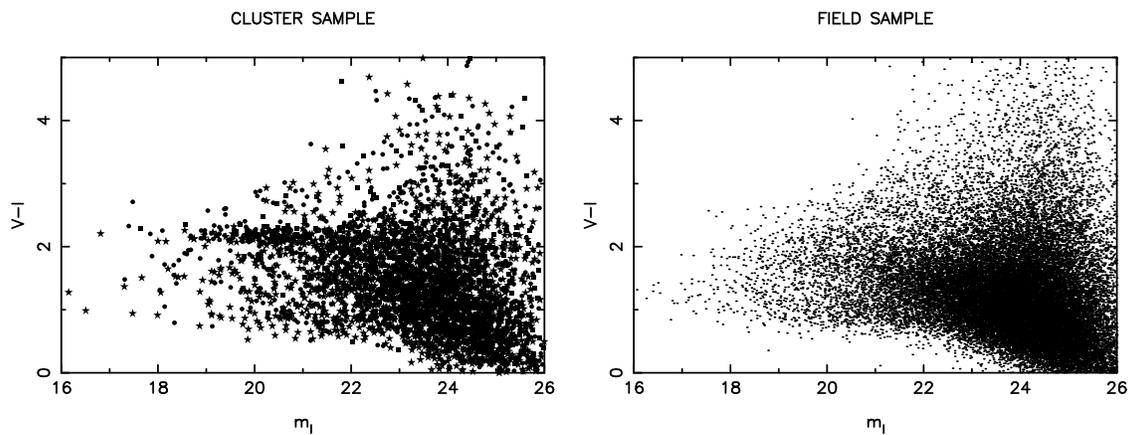,width=\figwidth,angle=-90}
\caption[Color-magnitude diagram]{Color-magnitude diagrams for
the cluster and field subsamples.
}
\label{fig:colmag}
\end{figure}

\begin{figure}[htbp!]
\centering\epsfig{file=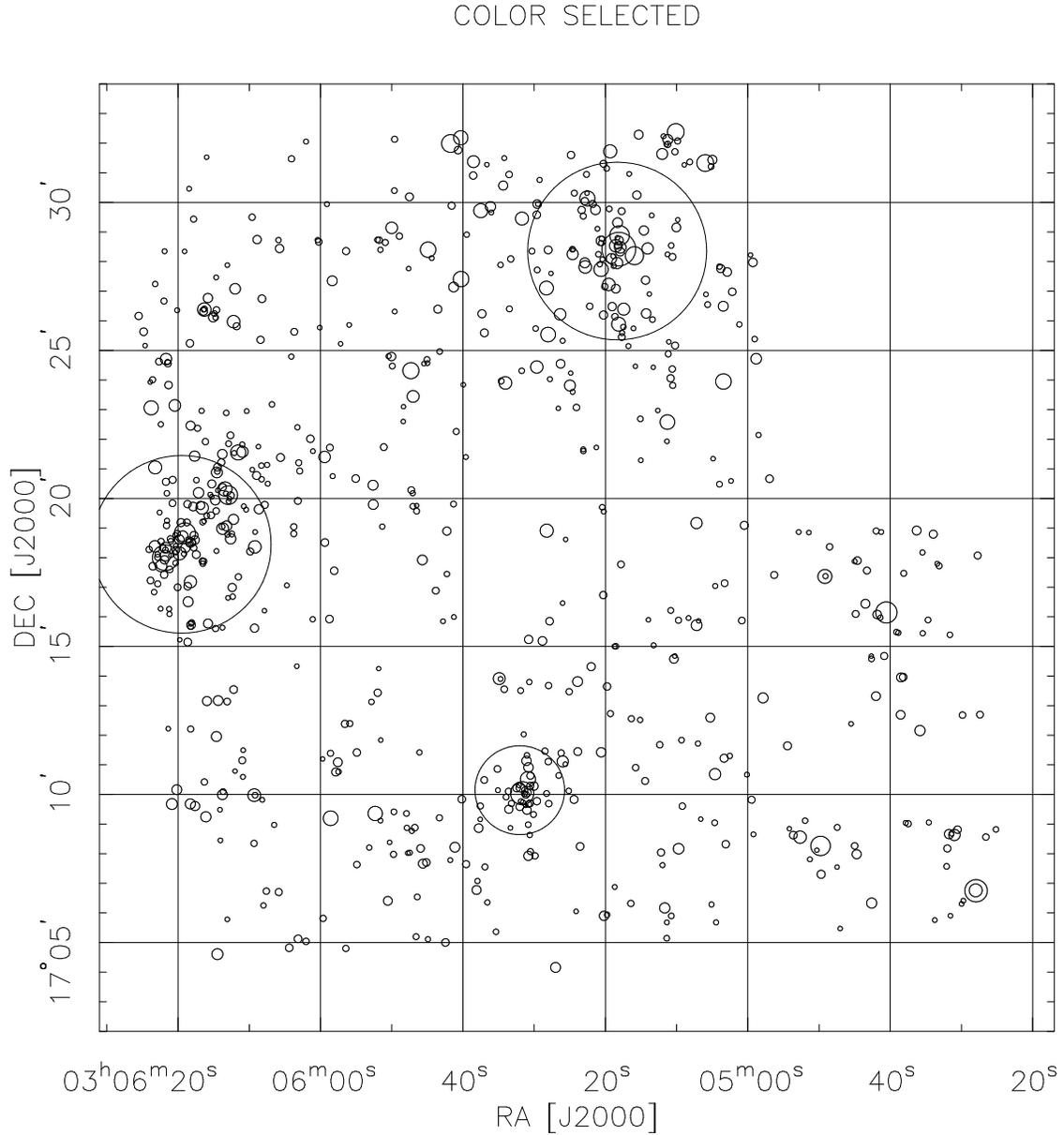,width=\figwidth,angle=-90}
\caption[Spatial distribution of color selected galaxies]{Early type galaxy sample.  
The spatial distribution of galaxies with colors $1.9 < V-I < 2.4$
and $I < 23$ chosen to select early type galaxies at around the redshift of the
super-cluster. The area of each circle is proportional to the flux. The range of colors
here corresponds to a depth $\Delta z \simeq 0.1$, so we would expect
to find a substantial contamination by `field' galaxies
in addition to the super-cluster galaxies.
}
\label{fig:colorselected}
\end{figure}

\begin{figure}[htbp!]
\centering\epsfig{file=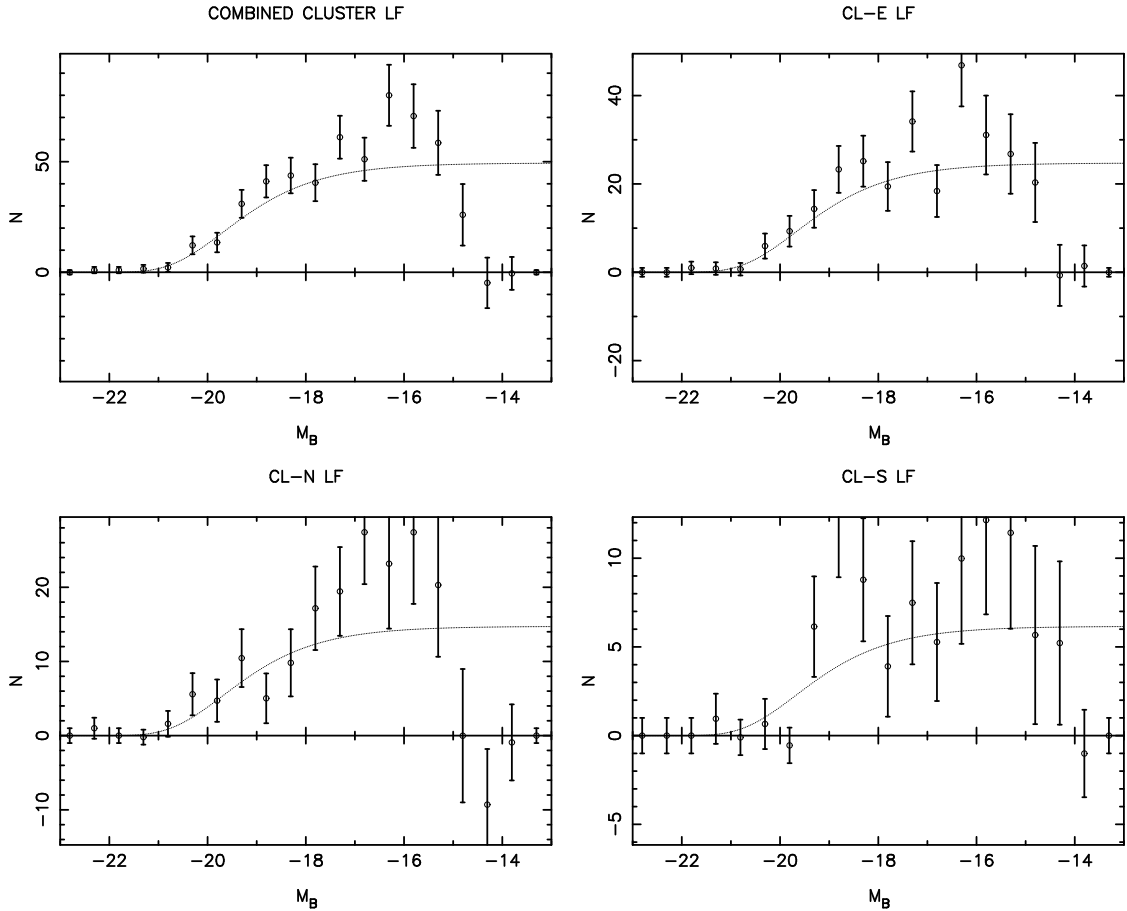,width=\figwidth,angle=-90}
\caption[Luminosity function]{Rest frame $B$-band luminosity functions
for the three X-ray clusters and for the three clusters combined.
The counts here have been corrected for field contamination
and the lines are fits to a Schechter function with $M_*$, $\alpha$
fixed at the values obtained for low redshift field samples.}
\label{fig:lumfun}
\end{figure}

The color magnitude diagrams shown in figure \ref{fig:colmag} 
reveal a distinct excess of galaxies in a narrow band of color around
$V - I = 2.2$ in the cluster regions, as predicted
for early type galaxies at $z=0.42$ from figure \ref{fig:galcolorvsz}.
By selecting only galaxies within a narrow strip in color-magnitude space
straddling this sequence
we can obtain fairly precise
estimates of the early-type luminosity functions for the three clusters, and for
the clusters taken together, since the fluctuations in the background
estimates
are greatly reduced.
The results are shown in figure \ref{fig:colorselected} where we plot the
spatial distribution of galaxies in a $\simeq 0.5$ mag wide strip 
(corresponding to a range of redshifts $\Delta z \simeq 0.1$)
and in
figure\ref{fig:lumfun} where we plot the
excess counts of the color selected galaxies (corrected for field contamination)
$\Delta N = A_c \times (n_c - n_f)$, where $A_c$ is the
area of the cluster circles, as a function
of $M_B = m_{I'} - c_{I'} + c_B + k_{BI'}(z = 0.42) - {\rm DM}(z)$.  
Also shown is the luminosity function of form
similar to those found by \citeN{eep88} and \citeN{lpem92}: 
$\Phi(L_B) = \Phi_* (L/L_*)^\alpha \exp(-L/L_*)$ with $M_* = -19.68$ and
$\alpha = -1.0$ and with 
amplitude $\Phi_*$ scaled to fit the
counts.  This standard, low-redshift, field galaxy luminosity function
seems to provide a reasonable fit to the cluster count excess down to
magnitudes well below the knee, and so integrating the luminosity functions
should give a good estimate of the total
red cluster light.  

To obtain estimates for the individual cluster luminosities we fit the
excess counts to the form above to determine $\Phi_*$ as
\begin{equation}
\Phi_* = \sum \Delta N f(M_B) / \sigma^2 / \sum f(M_B)^2 / \sigma^2
\end{equation}
where $f(M_B) = (L/L_*)^{1-\alpha} \exp(-L/L_*)$ and $\sigma$ is
the rms fluctuation in $\Delta N$ due to Poisson counting statistics.
We then integrate the luminosity functions to obtain the total
excess luminosities within the circles defining the cluster subsample.
These are tabulated in table \ref{tab:clusterlum}.  These luminosity functions
are only for the color selected subsample.  Performing the same exercise to
compute the total luminosity function using our synthetic $B$-band
magnitudes we find noisier results, but conclude that about 70\% of the
total excess light within our cluster apertures comes from early type galaxies,
in rough agreement with \citeN{fbm94} who found that about 25\% of the
galaxies in CL-N for instance were blue with `Balmer-line' spectra.

\input clusterlumtable.tex

The standard definition of cluster richness \cite{abell58} is the number of cluster galaxies 
in the range two magnitudes below the 
third brightest, with richness class
$R = 1$ corresponding to 50-80 galaxies etc.  
It is clear from inspection of figure \ref{fig:lumfun} that  these clusters are
not very rich by this measure.  Both CL-N, CL-E are borderline richness class 1
and CL-S is a very poor cluster.

The southern  cluster CL-S is particularly compact; most of the elliptical galaxies
lie within about $1'$ (or about $200 h^{-1}$kpc) of the X-ray
location.  This is almost an order of magnitude smaller than the
conventional Abell radius of $1500 h^{-1}$kpc.  This is consistent with the
compact X-ray appearance for this cluster, but also somewhat
surprising.  According to the standard hierarchical clustering picture,
clusters which turn around and virialize at time $t$ have
a density $\sim 170$ times
the critical density at that age: 
$\rho \sim 170 \rho_{\rm crit}(t) = 170 /(6 \pi G t^2)$.  
With $r \simeq
R_{\rm Abell}$ and $\sigma \simeq 1000$km/s, as typical of massive clusters,
we obtain a collapse time $t \simeq 10^{10} h^{-1}$yrs, similar to the
present age of the universe and therefore in reasonable agreement with the
theoretical expectation.  With $\sigma = 920$km/s and $r \simeq 200 h^{-1} $kpc
however,
we obtain a collapse time $t \simeq 2 \times 10^9$yr, which is much less than the
age of the universe at $z=0.42$.  If this cluster is virialized
and did indeed form by something
like the spherical collapse model then it did so a very long time ago at
$z \sim 3-5$ and
little visible matter has accreted onto it
since then.

%% file: clusterlumtable.tex
\begin{table*}[htbp!]
\begin{center}
\begin{tabular}{cccc}
\hline\hline
cluster & $\phi_*$ & $L/L_*$ & $h^2L/L_\odot$ \\
\hline
E+N+S & 59.27 & 128.70 & 1.49e+12 \\
CL-S & 6.34 & 13.77 & 1.60e+11 \\
CL-E & 30.30 & 65.81 & 7.63e+11 \\
CL-N & 17.87 & 38.81 & 4.50e+11 \\
\hline
\end{tabular}
\caption[Cluster luminosities]{Table of cluster luminosities.
}
\label{tab:clusterlum}
\end{center}
\end{table*}

%% file: lensing.tex
\section{Shear Analysis}
\label{sec:shearanalysis}

\subsection{Background Galaxy Selection and Shear Measurement}
\label{sec:galaxyselection}

We now describe how we select the faint galaxies which we
use for measuring the shear, how we correct for psf anisotropy and calibrate the effect of
seeing. We also describe how we construct a weighting scheme to
combine the signal from galaxies of a range of sizes and luminosities to
give a reconstruction with optimal signal to noise.

\begin{figure}[htbp!]
\centering\epsfig{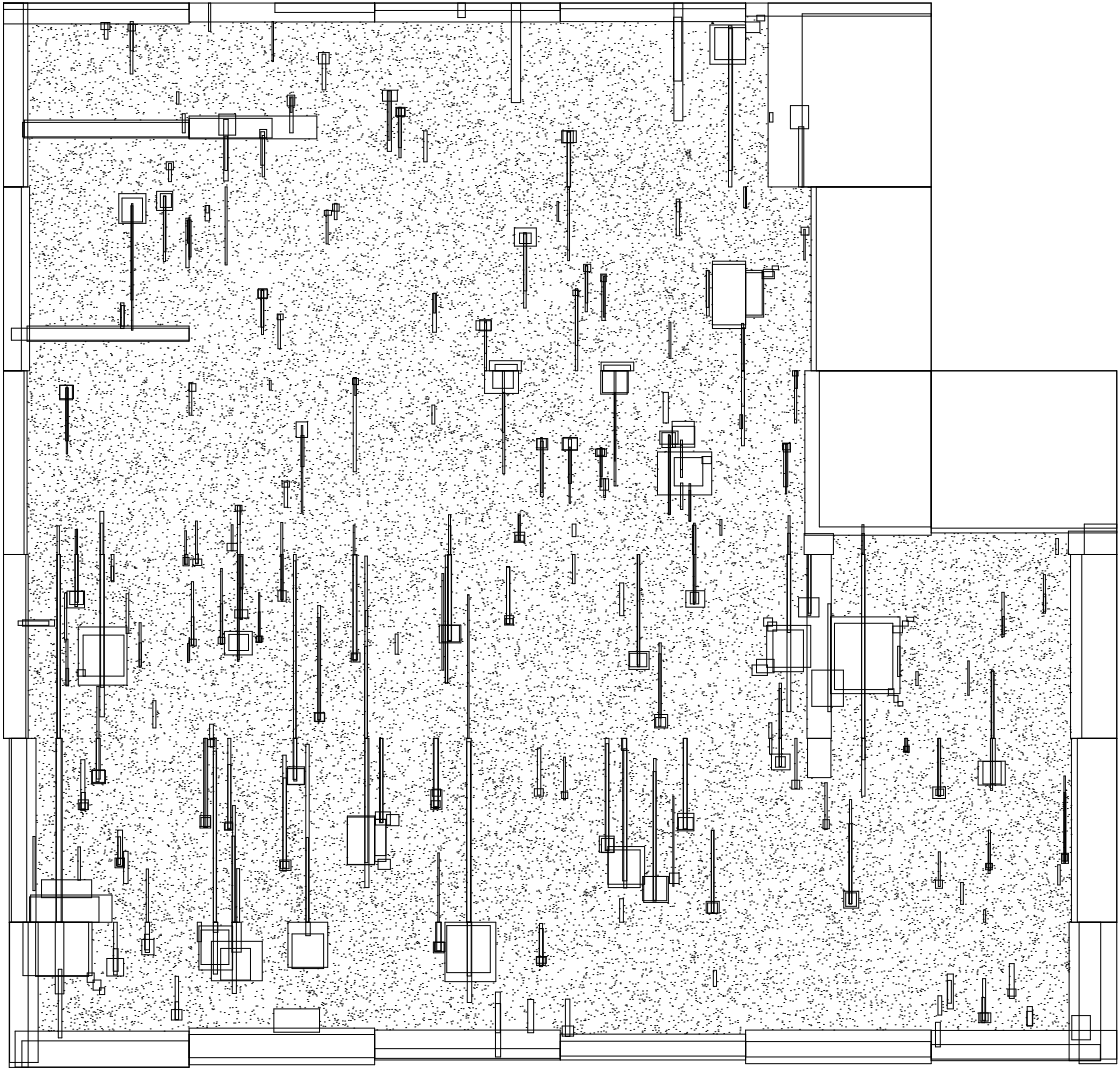}
\caption[Spatial distribution of faint galaxies sample.]
{Spatial distribution of the faint galaxy sample.  
They are clearly distributed very uniformly across the sky as
one would expect if they are very distant.  The rectangles were
placed to mask out diffraction spikes and trails from
bright stars.}
\label{fig:faintgals}
\end{figure}

Stars are easily visible to
$m_I \simeq 23$ (figure \ref{fig:sizemag}) and we remove these and also the
tail of faint objects with estimates half-light radii less than $0.3''$.
Some close pairs of galaxies get detected as
single objects, and these were filtered by applying a cut in ellipticity
(see below) which we set at $e = 0.7$. 
To obtain a sample which is distant and minimally contaminated by
cluster galaxies we make a cut at $\nu = 100$ to remove bright objects. 
The resulting catalog contains about $30,000$ objects, with 
a surface density of about $50$ per square arcmin.
The spatial distribution of this
faint galaxy sample is shown in figure
\ref{fig:faintgals}, which confirms that they are very uniformly distributed
across the field. 

The shear measurement is described in more detail elsewhere \cite{kaiser98}.  It is similar
in spirit to the \citeN{ksb95}  analysis in that we compute response functions
which tell us how the shape statistics respond to psf anisotropy and
to gravitational shear.  The fundamental shape statistic we use is the
flux normalized weighted second central moment of the measured
surface brightness $f(r)$:
\begin{equation}
q_{lm} = {\int d^2 r r_l r_m w(r) f(r) \over \int d^2 r f(r)} 
\end{equation}
where $w(r)$ is a Gaussian profile weight function with scale matched to
the size of the object.
KSB computed the response of 
the ellipticity or `polarization' 
defined as $e_\alpha \equiv M_{\alpha l m} q_{lm} / q_{pp}$,
where the pair of $2 \times 2$ matrices are 
$M_0 = \{\{1,0\},\{0,-1\}\}$,
$M_1 = \{\{0,1\},\{1,0\}\}$, 
to both psf anisotropies and, in the limit of well resolved
galaxies, to gravitational shear. 
\citeN{lk97} generalized this to obtain the post-seeing shear response function
$P_\gamma$ such that $\langle e_\alpha \rangle = \gamma_\alpha P_\gamma$
and which serves to calibrate the relation between the observable polarization
and the gravitational shear $\gamma_\alpha = (1/2) M_{\alpha lm} \Phi_{lm}$ where
$\Phi_{lm}$ is the projected gravitational potential.

In \citeN{lk97} the psf anisotropy was modeled as a low order polynomial
in position on the image.  Such a model is not adequate for observations
such as those considered here taken with a mosaic camera where one finds that
the psf varies smoothly across each chip, but changes discontinuously as
one passes across chip boundaries resulting in a complicated pattern
of psf anisotropies on the final averaged image. Since we know 
which source images contribute to a given galaxy image, 
we can compute an average of the psf anisotropy response function measured
from the individual images
modeling these as a smooth polynomial functions over each of the source images. 
This would
not be quite correct however, since the response function for the 
ellipticity, which is a non-linear function of the surface brightness,
is itself non-linear, so the averaged response function does not, in
general, correctly describe the response of an averaged galaxy image.
What we do instead is to compute the response functions
for the flux normalized $q_{lm}$ moments themselves.  Since
psf anisotropy does not affect
the net flux appearing in the denominator (we assume that our fluxes are
effectively total fluxes), the psf response is a linear
function of the sky brightness $f(r)$ and can therefore be averaged to
give the correct response for the $q_{lm}$ which we use to correct
the $q_{lm}$ values to what would have been measured by a telescope with a psf of 
same size and overall radial profile as the real psf but with
no anisotropy.
We then form a polarization $e_\alpha$ as above from the 
anisotropy-corrected
moments, and compute the post-seeing shear response function $P^\gamma$,
such that, on average, the polarization
$e_\alpha$ induced by a gravitational shear $\gamma_\alpha$ is
\begin{equation}
\langle e_\alpha \rangle = P^\gamma \gamma_\alpha
\end{equation}
We do not attempt to calibrate the galaxies individually; this tends to be
noisy and introduces non-linear effects.
Instead we split the galaxies up into a set of discrete
subsets by a coarse binning in the significance - half
light radius plane $(\nu, r_h)$, and compute an average
$P^\gamma$ for all the galaxies in each bin and thereby
compute shear estimates 
$\hat \gamma_\alpha = e_\alpha / \langle P_\gamma \rangle$.
Having split the galaxies up into classes by size and
magnitude in this way we can then compute an optimal weight as a function
of the bin.  To do this we assume that all of the
galaxies are at large, or at least similar, distance from the lens
and therefore feel essentially the same shear, and also that the
shear signal is much less that the random shear noise due to
intrinsic shape and measurement errors combined.  We compute the variance
in shear estimates 
$\langle \gamma_{\rm noise}^2 \rangle \simeq \langle e^2 \rangle /
\langle P_\gamma \rangle^2$ for each bin and assign weights inversely proportional
to the variance and such that the sum over galaxies of the weights is equal to
the number of galaxies (so the result is unbiased).
With this scheme, faint and/or
very small galaxies which are relatively noisy get assigned lower weights.

\begin{figure}[htbp!]
\centering\epsfig{file=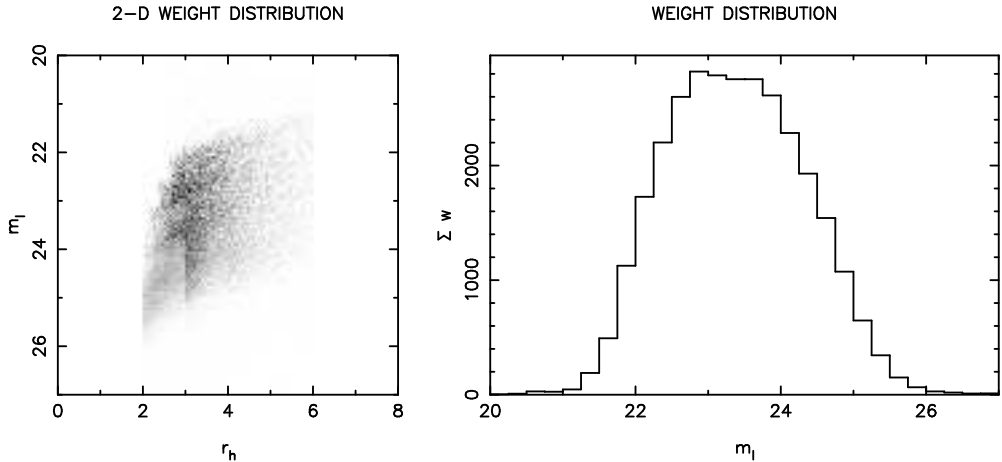,width={0.8 \linewidth},angle=-90}
\caption[Weight distribution.]
{Left panel shows the distribution of weight as a function of
size and magnitude.  The edges of the coarse bins 
can be seen as discontinuities
in the weight density. Right panel shows the distribution as a function
of magnitude.  Roughly half of our weight comes from galaxies brighter
than $m_I = 23.5$, where the redshift distribution is fairly well established
\cite{cshc96},
but half comes from fainter magnitudes where there is only very
poor sampling as yet.
}
\label{fig:weightdist}
\end{figure}

In figure \ref{fig:weightdist} we show the distribution 
of weight in the size magnitude plane
and we also show the distribution of weight as a function of
magnitude. 
The rms shear (per component) in the final catalog
is $\sigma_\gamma \simeq 0.32$ so  we should be able to measure shear to
a high precision. For instance, for measurement of the net shear
we expect the precision to be 
$\sigma_\gamma / \sqrt{N_{\rm gals}} \simeq 0.002$

\begin{figure}[htbp!]
\centering\epsfig{file=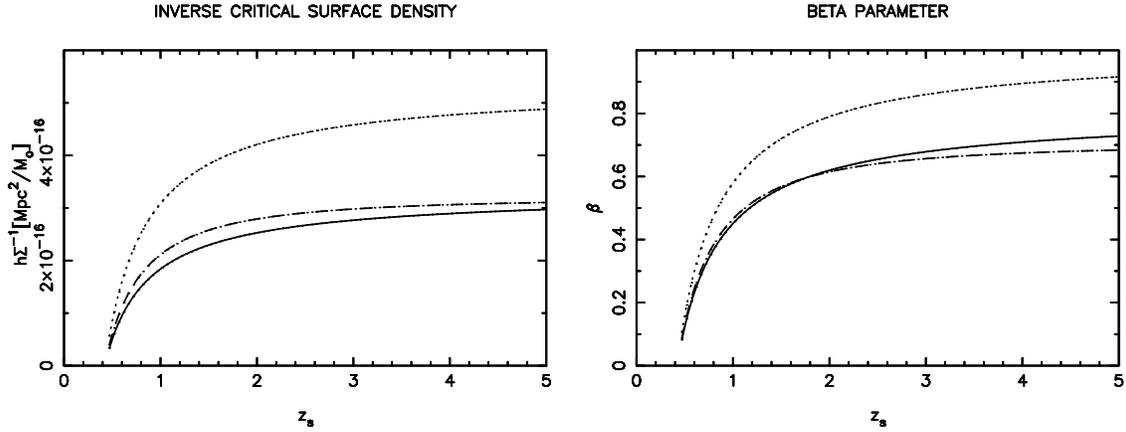,width=\figwidth,angle=-90}
\caption[Inverse critical surface density]
{Left panel shows the inverse critical surface density as a function of
source redshift for a lens at $z_l = 0.42$.  The physical surface density
of a lens is obtained by dividing $\kappa$ by a
weighted average inverse critical surface density taken over the distribution of
background galaxies.  The panel on the right shows the $\beta$ parameter;
for a lens of given velocity dispersion, X-ray temperature, the predicted shear is
proportional to $\beta$.  The three lines are for EdS (solid), empty open (dot-dash)
and $\Lambda$ dominated cosmology (dotted).
}
\label{fig:invsigmacrit}
\end{figure}

The largely unknown redshift distribution for galaxies at this
range of magnitudes and sizes is a major source of systematic
uncertainty in weak lensing mass estimation. 
For $m_I \lsim 23$ there are reasonably complete
statistical samples \cite{cshc96} but at fainter magnitudes things become
progressively uncertain.  
For sources at  a single redshift $z_s$, the dimensionless surface density $\kappa$
that we measure for a lens of physical mass surface density $\Sigma_m$ is 
$\kappa = \Sigma_m / \Sigma_{\rm crit}(z_s)$ where 
\begin{equation}
\Sigma_{\rm crit} = {c^2 \over 4 \pi G D_l} {D_s \over D_{ls}} \equiv \Sigma_{\rm crit}^\infty \beta^{-1}
\end{equation}
where $D_{ls}$ is the angular diameter distance relating
physical distance on the source plane to angles at the
lens and where the factor $\beta \equiv D_{ls} / D_s$ is the distortion
strength relative to that of sources at infinite distance.
With sources at a range of distances what we measure is the physical
surface density times an average inverse critical density, or
$\kappa = (\Sigma_m / \Sigma_{\rm crit}^\infty) \langle \beta \rangle$.
The critical surface density and the $\beta$ parameter are shown, for
a lens redshift $z_l = 0.42$ in figure \ref{fig:invsigmacrit}.

Using the photometric redshifts in the Hubble Deep Field
from \cite{sly97} and assuming
an empty open universe
gives $\langle \beta \rangle \simeq 0.46$. 
Using an alternative, and more recent, HDF 
photo-z catalog calculated \cite{fly98}  we find 
similar results. Both catalogs contain about 1000 galaxies. 
The $\langle \beta \rangle$ value was calculated by transforming 
HDF $I_{\rm F814W}$ AB magnitudes to Cousins $I$-band magnitudes 
according to the WFPC2 Handbook and HDF web-page, and corresponds to that for a 
single screen of background galaxies at $z_s \simeq 1.0$.  

There is considerable uncertainty in this value.  There are some 
questions about the
reliability of photometric redshifts, the calculation above does not
incorporate the weighting as a function of size, and the cosmological
background model may be inappropriate. Another result 
which casts some doubt on the
detailed accuracy of the distribution of photometric redshifts
is that they are somewhat hard to reconcile with weak lensing
results for $z \simeq 1$ clusters (\citeNP{lk97}; \citeNP{cklhg98}).
In what follows we shall
adopt a value of $\Sigma_{\rm crit} = 4.33\times 10^{15} h  M_\odot / {\rm Mpc}^3$, slightly higher than that derived from the photometric redshifts, and
corresponding to a single screen redshift of
$z_s \simeq 1.5,\; 1.2; 0.75$ for EdS, empty open and $\Lambda$ dominated cosmologies
respectively.  

\subsection{Predicted Dimensionless Surface Density}
\label{sec:kappapred}

We now compute the dimensionless surface density
$\kappa$ that would be generated by the 
optical structures assuming
that galaxies are good tracers of the total mass with some 
constant mass-to-light ratio 
We know of course that
on mass scales of the order of tens of kpc the dark matter 
distribution is much more extended than the light, but on larger
scales this `what you see is what you get' picture remains an interesting and
viable hypothesis.

If all of the structures lay at the same lens-redshift $z_l$ this would be
straightforward since we then have
\begin{equation}
\kappa \equiv {\Sigma_\rho \over \Sigma_{\rm crit}}
= (M/L) \Sigma_l / \Sigma_{\rm crit}(z_l)
\end{equation}
where $\Sigma_l$ is the rest frame luminosity
surface density, which we can compute from the observed surface
brightness.

\begin{figure}[htbp!]
\centering\epsfig{file=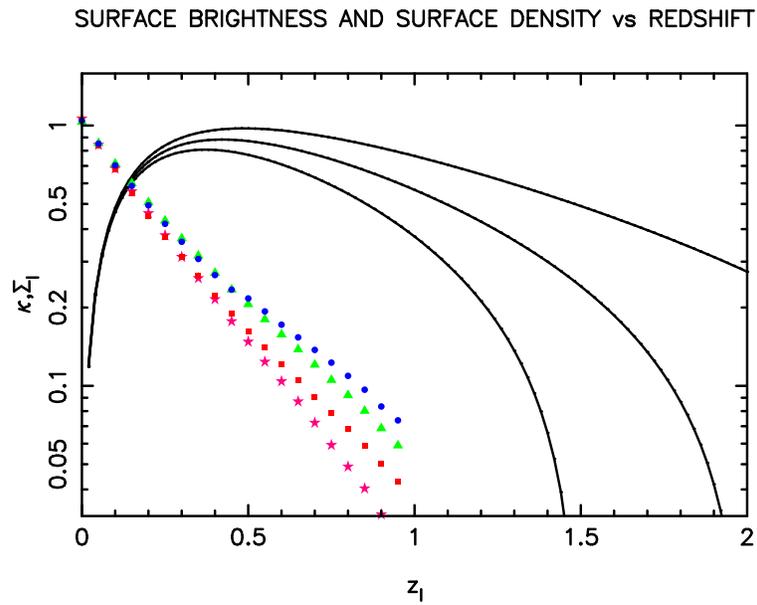,width={0.6 \linewidth},angle=-90}
\caption[Surface brightness and surface density vs lens redshift.]
{The filled symbols show the observer-frame surface brightness
(in the UH8K $I$-band) of a cluster lens as
a function of lens redshift and for various galaxy types, assuming no
evolution in the rest-frame surface brightness. 
Symbols encode the galaxy type as follows:
E0 = stars; Scd = squares; Sbc = triangles; Im = circles.
The lines show the inverse critical surface density, 
which for a lens of a given
physical surface density density, 
is proportional to the observed dimensionless
surface density $\kappa$, for sources at redshifts $z_s = (1.5, 2.0, 3.0)$.
}
\label{fig:kappasigma}
\end{figure}

This would be adequate to compute the contribution just from the $z=0.42$ supercluster,
but, as we have seen, there are other structures at different redshifts that we would like to
include, and 
one cannot then use this simple procedure because it
would cause one to grossly overestimate the effect of
foreground structures.  This is because, for a lens with
given rest-frame surface luminosity and mass surface density,
the observed surface brightness falls very rapidly with increasing lens
redshift (roughly as $(1+z)^{-4}$ but with an extra decrease for
early type galaxies which tend to have $\lambda f_\lambda$ increasing with $\lambda$)
while the dimensionless surface density $\kappa \propto 1 / \Sigma_{\rm crit}$
is a strongly 
increasing function of lens redshift for $z_l \ll z_s$.  These trends are
quantified in figure \ref{fig:kappasigma}.  For early type galaxies,
and in the $I$-band, the ratio of $\Sigma_l / \kappa$  falls
by about a factor of 3 as the lens redshift is varied from
0.2 to 0.4.
Note also that high redshift clusters,
if they exist, could give quite a strong lensing effect 
while being relatively inconspicuous in photometric surveys (the $\kappa$ 
curve is rather flat, while the $ \Sigma_{l, {\rm obs}}$ curve
plunges rapidly).  For example, a cluster at redshift unity produces
a shear per surface brightness about one order of magnitude larger than 
the same cluster at redshift $0.42$ (this is neglecting the evolution of the
cluster galaxies which tends to act in the opposite direction).

To make an accurate prediction for the dimensionless surface density
$\kappa$ then, it is not sufficient just to use the net surface brightness;
one needs some idea of the distances to the galaxies.
Ideally, one would obtain redshifts for all of the bright galaxies
and then one could correctly account for this, but as yet we do not
have this information.  A solution to this problem is suggested by
figure \ref{fig:galcolorvsz}. If we only use galaxies redder than 
$V-I \simeq 1.6$ then we should see elliptical galaxies at
$z \gsim 0.2$, and later type galaxies only at $z \gsim 0.4$.
For an elliptical galaxy, we can read off the redshift 
from the color using figure \ref{fig:galcolorvsz} (we model the
color-redshift relation as a linear trend fixed by the values
at $z = 0.2$ and $z = 0.4$).  
We can then compute
the physical luminosity, and thereby the mass (assuming a constant
$M/L$), and thereby compute it's contribution to the $\kappa$ field.
For a single galaxy with observed magnitude
$m$, and for an EdS cosmology,
the contribution to an image of $\kappa$ from a galaxy which falls 
in a pixel with solid angle $d\Omega$ is
\begin{equation}
\label{eq:kappadomega}
\kappa d\Omega = {M \over L}
{4 \pi G M_\odot a_0 \over c^2 (10{\rm pc})^2}
(1+z)^3 {\omega_l \omega_{ls} \over \omega_s}
10^{0.4 (M_{B\odot} - m - k(z))}
\end{equation}
where $M/L$ is in solar units and $\omega = 1 = 1 / \sqrt{1 + z_l}$.

If we generate a surface density prediction $\kappa$ in this manner
we will, to a very good approximation have isolated the
contribution from the early type galaxies at $z \gsim 0.2$ alone.  
To see why, consider first the
blue galaxies associated with the clusters at $z \simeq 0.42$, which
we have seen contribute at most about 30\% of the cluster $B$-band light.
Some fraction of these --- the very reddest --- will survive
the color limit of $V-I = 1.6$, but by the above prescription they
will be assigned a redshift of $\sim 0.2$, rather than their actual
redshift of $0.42$ and so their contribution to $\kappa$ 
as calculated above will be
suppressed in proportion to the lens redshift dependent factors in equation
(\ref{eq:kappadomega}):
$(1+z)^3 \omega_l(z_l) \omega_{ls}(z_l, z_s) 10^{-0.4 k(z_l)}$
or about a factor 6 in this example, so,
even if we assume that say 30\% survive the color selection, the
contribution to $\kappa$ will only be about $1/20$ of the true contribution
of all the blue galaxies, or about $1/60$ of the contribution
of the elliptical galaxies.  At higher redshift
a larger fraction of later type galaxies survive the
color selection, but
for these the contribution to $\kappa$ will be suppressed by
an even larger factor.  Moreover, these galaxies are found to be very
nearly uniformly distributed on the sky; thus their contribution
to the {\sl fluctuating\/} part of $\kappa$ --- which is 
entirely responsible for the
shear we are trying to predict --- 
is expected to be very small and it should be safe to
neglect them as compared to the early type galaxies at 
lower redshift. Also, our linear extrapolation of the color-redshift
relation exceeds the actual predicted colors for $z \gsim 0.6$, and this effectively
suppresses the contribution from higher redshifts.

With this color-based redshift estimation we
obtain a prediction for $\kappa$ which will be correct if
early type galaxies trace the mass with some constant $M/L$,
(though missing some or all of the contribution from very distant
structures where our selection function cuts off around $L_*$
or brighter).
We should stress that this is a rather unconventional
hypothesis.  The general picture of the structure of superclusters,
supported by the \citeN{dressler80} morphology density relation, 
galaxy correlation functions \cite{lmep95} and
studies of individual structures like the Virgo Supercluster, 
is of the elliptical galaxies residing in dense compact clusters,
with the later type galaxies (which dominate the total luminosity)  being much more extended.
If this supercluster conforms to the norm, then the mass
should be more extended than the early type galaxies and we should see
this in the gravitational shear field, and if one
incorporates statistical `biasing' (e.g.~\citeNP{bbks86}) then the mass would be expected
to be still more extended. 

\begin{figure}[htbp!]
\centering\epsfig{file=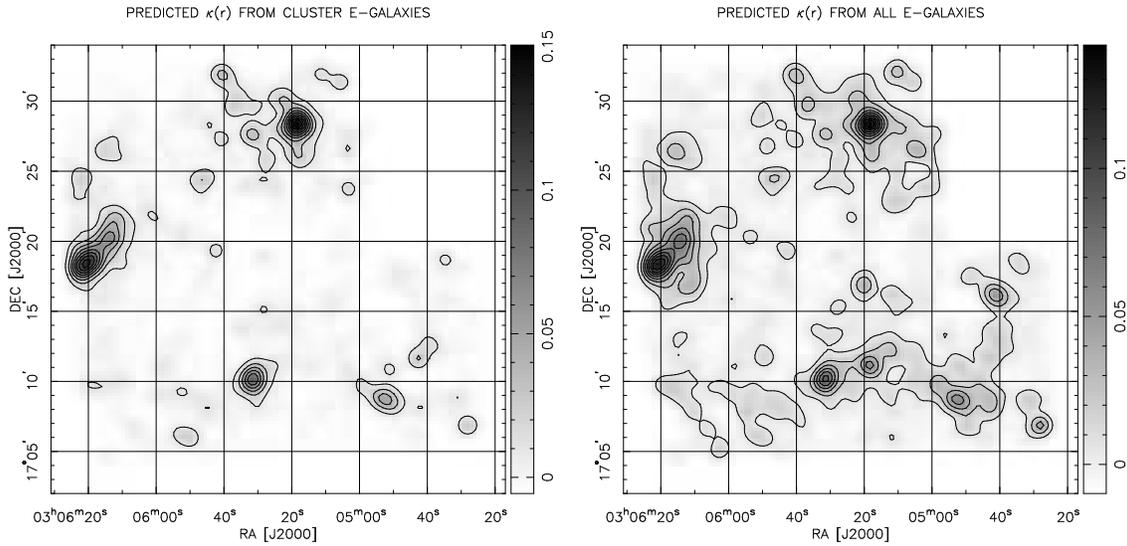,width=\figwidth,angle=-90}
\caption[Predicted surface density.]{Predicted dimensionless surface density $\kappa$. 
Images have been smoothed with a $31''$ arcmin Gaussian low pass filter.
The mean has been subtracted from each image.
The left panel shows the predicted $\kappa$ only from galaxies in a narrow band of
color around $V-I \simeq 2.2$ for elliptical galaxies
$z = 0.42$.  The right hand panel
shows the results for all galaxies redder than $V-I = 1.6$ and with
contribution given by equation \ref{eq:kappadomega}. 
}
\label{fig:lightmaps}
\end{figure}

The resulting predicted $\kappa$ field smoothed with a 
Gaussian kernel of scale length $31''$ is shown in figure \ref{fig:lightmaps}.
We show both the $\kappa$ just due to the $z \simeq 0.42$ cluster galaxies and
also for a broader color selection. 
These were made with $M/L = 200h$ which, as we shall see, is similar to the
value preferred by our shear measurements.
The differences
are not large, though there is some contribution from the structure
extending West from CL-S which lie at $z\simeq 0.3$, and there are
several other weaker features.

The luminosity density field we measure is obtained from 
fluxes measured within apertures around the galaxies (these were taken to be three times the
Gaussian scale length of the detection filter) and so will differ slightly
from the total luminosity dependence because of two competing effects;
the apertures are finite and so miss some of the light, 
but also sometimes overlap so that light gets double counted.  To test how large
the net bias is we have made estimates of the total extra-galactic 
surface brightness field
in two ways: by summing the fluxes from the catalog and by using all the
light in an image from which we have masked out all the stars.
The results agree to better than 10 percent, so we conclude that
we are measuring essentially the total luminosity.

\subsection{Predicted and Observed Shear}
\label{sec:shearanal}

Armed with the predicted $\kappa(r)$ from the previous section we can compute the shear field and
compare with what we observe.
To generate the predicted shear from the $\kappa$ images we
solve  the 2-dimensional Poisson's equation by FFT techniques
(padding the predicted $\kappa$ image out to twice its original size with
zeros to suppress the effect of periodic boundary conditions) to obtain the
surface potential $\Phi$, and
then differentiate this  to compute the shear $\gamma_\alpha = M_{\alpha l m} \Phi_{lm} / 2$.
This provides us with an image of the shear field, and with this we can generate for
each galaxy in our faint galaxy catalog the value of the predicted shear, and we can then
ask, for instance, what is the multiple of the predicted shear which, when
subtracted from the observed shear minimizes the summed squared residuals. We can
also bin the galaxies according to predicted shear --- here we treat the two
components of the shear for each galaxy as independent measurements ---  and plot the average measured
shear versus that predicted.
To minimize the effect of shear generated by structures outside the
field in this analysis, we subtract, from both the observed and predicted shears, the
mean shear.

\begin{figure}[htbp!]
\centering\epsfig{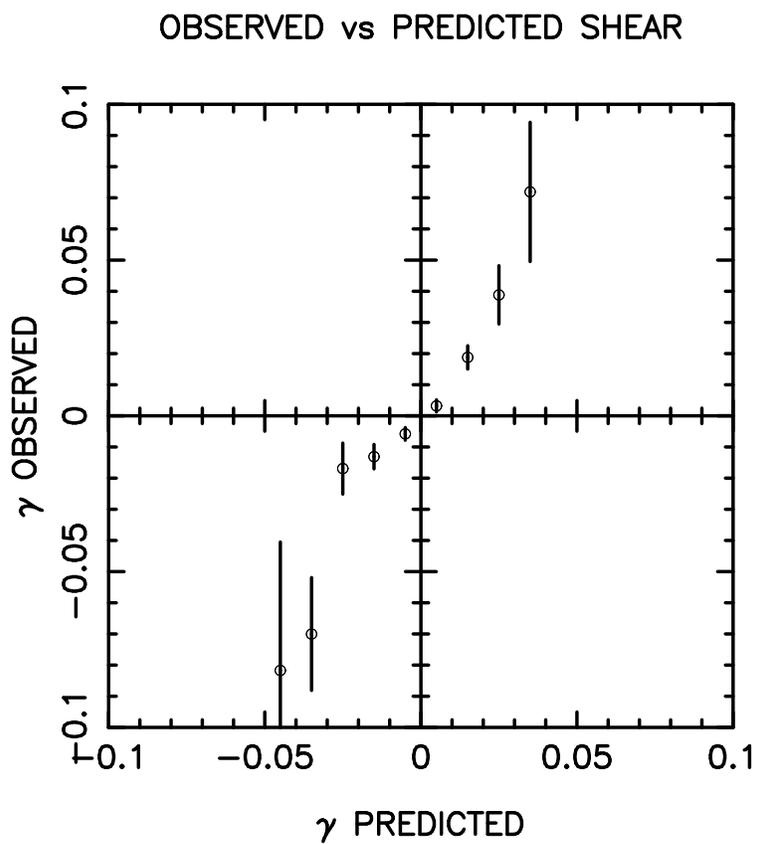}
\caption[Measured and predicted shear]{Predicted vs observed shear assuming $M/L_B = 200h$.
Error estimates were obtained from an ensemble of catalogs with  randomized shear values.
}
\label{fig:shearshearplot}
\end{figure}

If we assume that light traces mass and that the errors are predominantly
in the shear measurement then to obtain the mass-to-light ratio we
should compute $\alpha = \sum \gamma^{\rm obs} \cdot  \gamma^{\rm pred} / 
\sum \gamma^{\rm pred} \cdot  \gamma^{\rm pred}$, and then multiply this by our nominal
value to obtain $M/L_B = 200h \alpha$.  We find $\alpha = 1.20$, with
this value giving a reduction in 
$\chi^2 \equiv \sum (\gamma - \gamma_{\rm model})^2 / \sigma_\gamma^2$ 
of $79$ relative to the value for $\alpha = 0$, from which we
infer $M/L_B \simeq 240 h$.
The correlation between the predicted and observed shear
is shown in figure \ref{fig:shearshearplot} and is significant at about the $\sim 10$-sigma level.

A weakness of this method is that the light was smoothed, whereas the shear
values are unsmoothed.  If light really traces mass fairly on all scales,
then this would cause our estimate to underestimate the
true mass to light ratio by a factor
$\int d^2k P(k) \omega(k)^2 / \int d^2k P(k) \omega(k)$ where
$\omega(k)$ is the smoothing filter
and $P(\omega)$ is the luminosity power spectrum.  If mass were distributed around galaxies
in halos of shape proportional to $\omega(r)$, however, then we would
obtain the correct $M/L$. Conversely, if it were the case that the dark matter
distribution is more extended than our smoothing filter then we will
have underestimated the global $M/L$.
We will return to this issue shortly.

\subsection{2-D Surface Density Reconstruction}
\label{sec:reconstruction}

We now show 2-D mass-reconstructions 
and compare with the projected luminosity density.  We will also
compare the results obtained using the $V$- and $I$-band observations
separately, and with the level of noise expected due to
the random intrinsic galaxy shapes.
We use the \citeN{ks93} reconstruction algorithm. This is
a stable and fast reconstruction method which has very simply
defined noise properties; essentially Gaussian white
noise.  Its main drawbacks are
that it does not properly account for patchy data and the finite
data boundaries.  We have found from simulations that the former 
is not a serious problem for these data, but the
latter is worrying, particularly for cluster CL-E, which lies right
at the edge of the field and so some of the shear signal is lost, resulting
in a suppression of the recovered mass.  
To get around this we proceed as  follows:
first we make a shear field image prediction from the 
constant $M/L$ $\kappa$ prediction.  We then
sample this at the actual positions of our faint galaxies to generate
a synthetic catalog (that which would have been observed with no intrinsic
random shape or measurement noise), and
then generate a reconstruction from that synthetic catalog which will
have the same finite-field bias as the real reconstructions. 
To match the spatial resolution to that of the real reconstructions 
(a $45''$ Gaussian) we make the predicted shear with a smoothing 
scale $45'' / \sqrt{2}$ and make the reconstruction from the
synthetic catalog with the same smoothing. 
Note that while correctly accounting for the finite field effect on structures
within the field, the actual shear may still feel some effect from
structures outside of the field.

\begin{figure}[htbp!]
\centering\epsfig{file=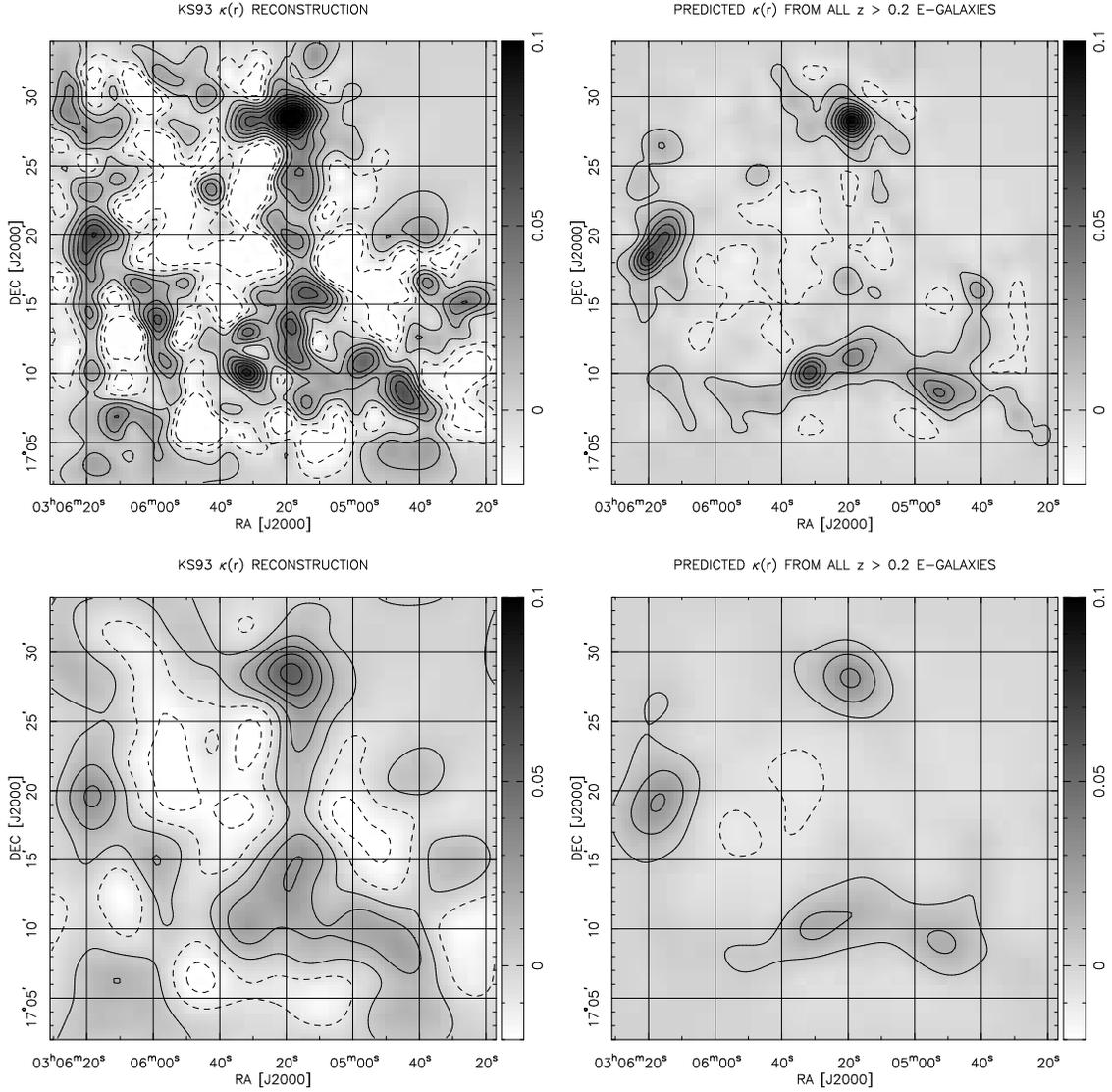,width=\figwidth,angle=-90}
\caption[Mass and light distributions.]
{Left panel shows the recovered mass using the \citeN{ks93} reconstruction method.
Right panel shows the predicted 2-D reconstruction using
the shear predicted from the light-map (but sampled at the actual
locations of the faint galaxies) which should have the
same finite field biases as the real reconstructions. The lower panels
show the same images smoothed with a Gaussian of scale $90''$ }
\label{fig:massvslight}
\end{figure}

In figure \ref{fig:massvslight} we show the mass reconstruction from
our combined faint galaxy sample on the left and that
predicted (for a nominal $M/L_B = 200h$) on the right.  There is generally
quite good agreement between the mass and the light.  All three of the
main $z \simeq 0.42$ clusters are clearly detected in the
reconstruction, and the locations of these peaks coincide very accurately.
We see no obvious indication that the mass in these clusters is
more extended than the elliptical galaxy light.
The lower redshift structure extending to the South-West of CL-S is 
also recovered in our reconstructions.
There is some indication of a `bridge' linking CL-S, CL-N,
which is suggestive of the kind of features 
predicted by the `cosmic-web' theory \cite{bkp96}.
Part of this bridge is a clump at 
$\alpha \simeq 3^h05^m15''$, $\delta \simeq 17^\circ16'$ 
which roughly matches a predicted, though somewhat weaker, feature.
However, this is not at the supercluster redshift, and so part of
the `bridge' may be a coincidental projection.
There is a rather conspicuous mass clump around
$\alpha \simeq 3^h05^m57''$, $\delta \simeq 17^\circ14'$ 
which has no apparent counterpart in the predicted field.

There is considerable noise in the reconstructions.  The 
noise power spectrum is white, while the galaxy power spectrum is
relatively red, so we might expect to see better agreement if
we further smooth both the light and mass maps, and to this end we
show in
the lower pair of panels of figure \ref{fig:massvslight} the
results for a smoothing scale of $\simeq 90''$.

\begin{figure}[htbp!]
\centering\epsfig{file=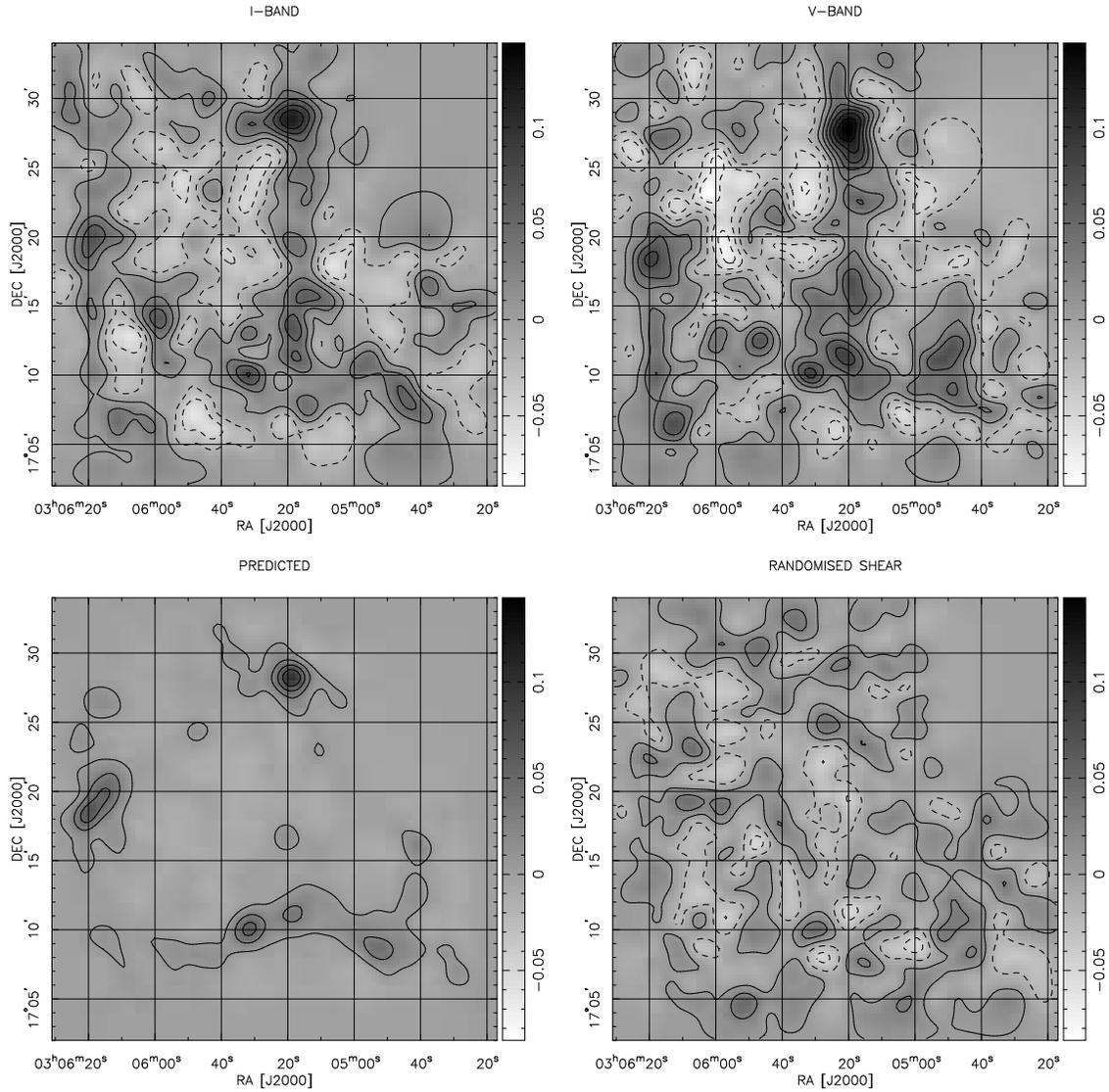,width=\figwidth,angle=-90}
\caption[Comparison of mass reconstructions.]{Upper panels
show the reconstructions from catalogs made from the $I$ and $V$ band
images separately and show generally very good agreement.  Lower left
panel shows the early-galaxy $M/L = 200h$ prediction. Lower right panel
shows the reconstruction from a catalog with randomized ellipticities,
which indicate the expected noise fluctuations due to intrinsic random galaxy shapes.}
\label{fig:massmaps}
\end{figure}

Figure \ref{fig:massmaps} shows
mass reconstructions made from the I-band and V-band observations
separately. 
These show generally very good agreement, with most of the
features described above being visible in both images. 
The anomalous `dark clump' lying between CL-S and CL-E is however not very evident in
the $V$-band reconstruction.
Also shown is a reconstruction
obtained using a catalog with the real faint galaxy positions but with
the measured shear values shuffled (i.e.~reassigned to the
galaxies in random order).  
The fluctuations in this
random reconstruction show the expected noise level due to
random intrinsic ellipticities. 
We have made an ensemble of 32 such realizations and
we use these below to quantify the significance of our results.

\subsection{Cluster Masses and Mass to Light Ratio's}
\label{sec:clusterml}

We now estimate the masses of the three $z \simeq 0.42$ clusters.
We will use these to derive mass-to-light ratios, and also to compare with
independent mass estimates from the X-ray temperature and velocity
dispersions. We compute the masses within circular apertures
of radius $1'.5$ and $3'.0$ from the reconstructions (which have been
smoothed with a $45''$ Gaussian), and we estimate mass-to-light
ratios by dividing the mass by the predicted mass within
the same apertures.  Thus, the $M/L$'s we obtain should be
corrected for finite field effects. The results are tabulated in
table \ref{tab:clusterml}.

\input clusterml

Table \ref{tab:clusterml} shows an increase 
in $M/L$ with aperture radius for all three clusters.  For CL-E,
CL-S, this is slight and not particularly significant.  For CL-N the
increase is larger, but we caution that there is a
foreground cluster which lies just to the East of CL-N and which falls
within the larger aperture.  While we have argued above that there should
in general be a rather weak contribution from low redshift structures,
in this case the foreground cluster is particularly bright, and so may contribute
significantly to the increase in $M/L$ for CL-N.  Redshift information
is needed to properly quantify this.  We note that these $M/L$ values,
averaging around $260h$ for our $1'.5$ apertures (which give the
highest signal to noise for $M/L$) are much lower than has been found for
other, generally more massive, clusters like A1689 \cite{ft97}.  
In so far as only a
tiny fraction of the Universe resides in such super-rich systems, it may well
be that the value obtained here is more representative.
It is also interesting to note that these $M/L$'s only include the
contribution from early type galaxy light in a rather narrow band of
color.  CL-N was found by \citeN{fbm94} to contain a number of
bluer `Balmer-line' galaxies, about $\sim 25$\% of the total, which
would bring the total
$M/L$ is more in line with that for CL-S, which they found to be almost
entirely early galaxy dominated.

It is of interest to compare these masses with those obtained from the
CL-S temperature and with velocity dispersions for all 3 clusters.
For CL-S, our mass is in  good agreement with the X-ray mass
(which in turn is in good agreement with the velocity dispersion).
For CL-N,  
we model the galaxies as an isothermal gas with profile
index $d \log n / d \log r \simeq -1.5$ as measured from the  observed
counts profile to find a predicted $\overline{\kappa} \simeq 0.93$ (for
the $1'.5$ aperture) as compared to the measured value
$\overline{\kappa} = 0.076 \pm 0.01$, again in reasonably good
agreement given the statistical and systematic modeling uncertainties. 
For CL-E the same exercise gives a predicted 
surface density about a factor two higher than what we measure,
but this may be due to departures from assumed equilibrium in this
clearly non-relaxed system.

\subsection{Light-Mass Cross-Correlation}
\label{sec:crosscorrelation}

In this section we show the results of cross-correlating the
mass and the light. Our goals are both to determine the $M/L$
parameter and also to test the hypothesis of a constant $M/L$
which is independent of scale.
We perform the correlation both in real space and Fourier
space.

\begin{figure}[htbp!]
\centering\epsfig{file=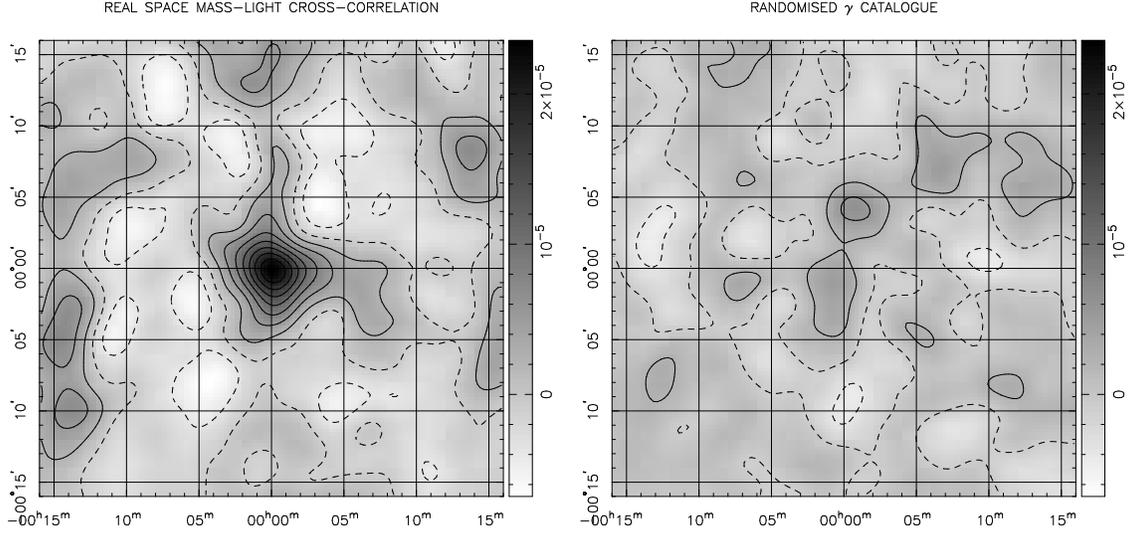,width=\figwidth,angle=-90}
\caption[Mass-light cross-correlation.]{Cross-correlation of light with mass reconstruction and
with randomized catalog reconstruction. Left panel shows that actual cross
correlation and right panel shows the result of cross correlating the light
with a $\kappa$ reconstruction generated from a catalog with
randomized (shuffled) shear values.
}
\label{fig:ccf}
\end{figure}

The real space cross-correlation function is shown in
figure \ref{fig:ccf}. In computing this we padded the source images
with zeros to twice the original image size.
Also shown is a typical random realization from the ensemble
used to estimate the noise fluctuations.
We see a strong cross-correlation peak at close to zero lag.
The significance, estimated as the strength of the zero-lag
correlation relative to the rms found from our ensemble of
randomized catalogs
is $9.3$-sigma. The correlation strength 
at zero lag implies a mass-to-light
ratio $M/L = (224\pm24)h$ in solar units.

\begin{figure}[htbp!]
\centering\epsfig{file=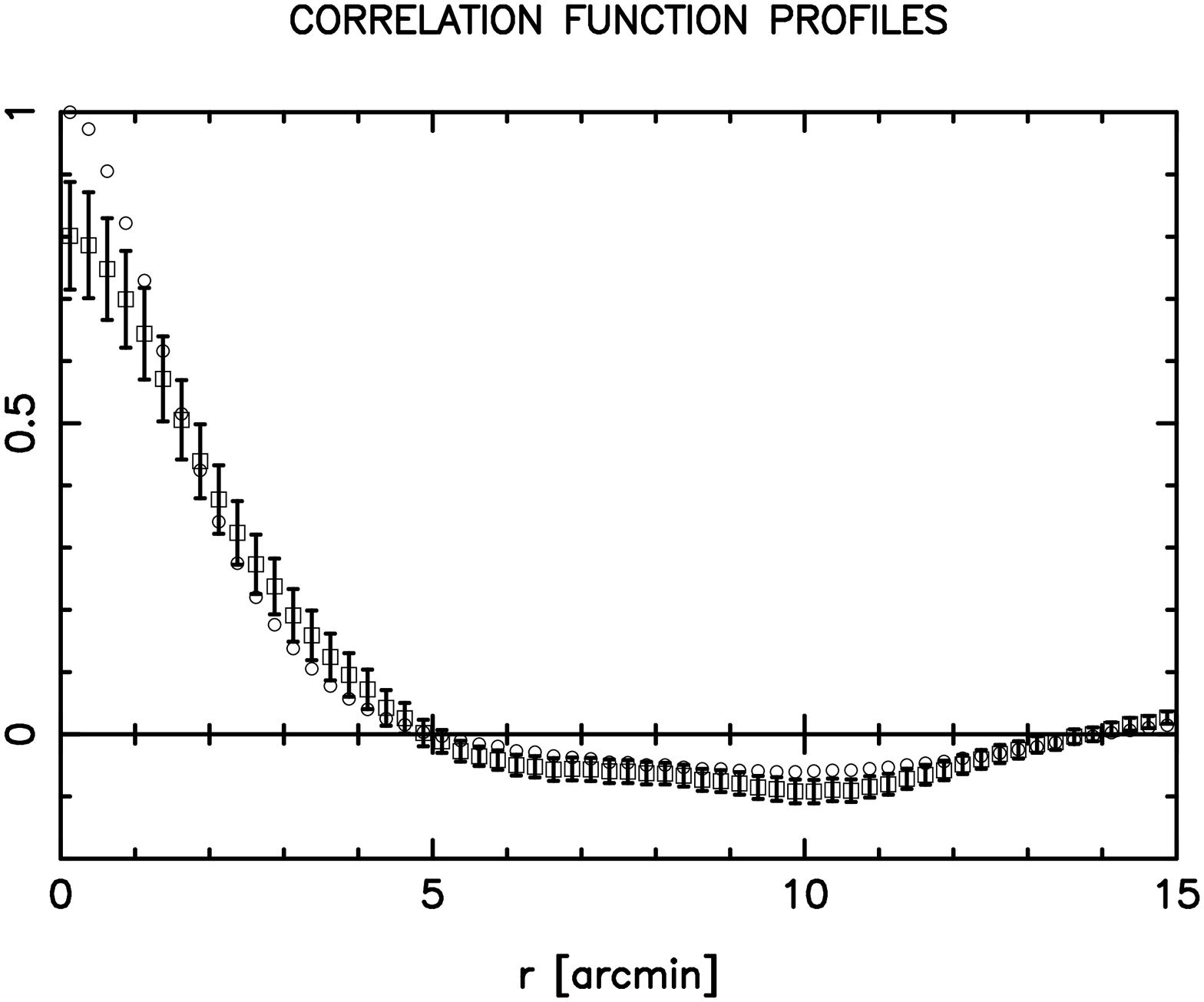,width=\figwidth,angle=-90}
\caption[Mass-light cross-correlation profile.]{Azimuthally averaged profile of the
mass-luminosity cross-correlation function (boxes with error bars) and
the luminosity auto-correlation function (circles). In normalizing these we have adopted
a mass-to light ratio of $M/L = 280h$, somewhat larger than that obtained by
simply comparing the correlation strengths at zero lag.
}
\label{fig:ccfprofile}
\end{figure}

To see if $M/L$ changes with scale, 
in figure \ref{fig:ccfprofile} we show the profile of the luminosity-mass cross-correlation and
the luminosity auto-correlation function. These are very similar in shape.
We caution that these should in no way be taken to be indicative of the
mass-luminosity auto-correlation function in general; the field is small, so the
`cosmic-variance' or sampling variance is very large, and the field is also
unusual in that it was chosen because it contains a prominent supercluster.  What this figure
shows though is that aside from a slight enhancement of the luminosity
auto-correlation function at very small lag --- which might plausibly be due to
dynamical segregation --- the cross- and auto-correlation functions have very
similar shapes indeed.  This suggests that on scales $\gsim 1'$ the early type galaxies 
trace the mass quite faithfully.  

\begin{figure}[htbp!]
\centering\epsfig{file=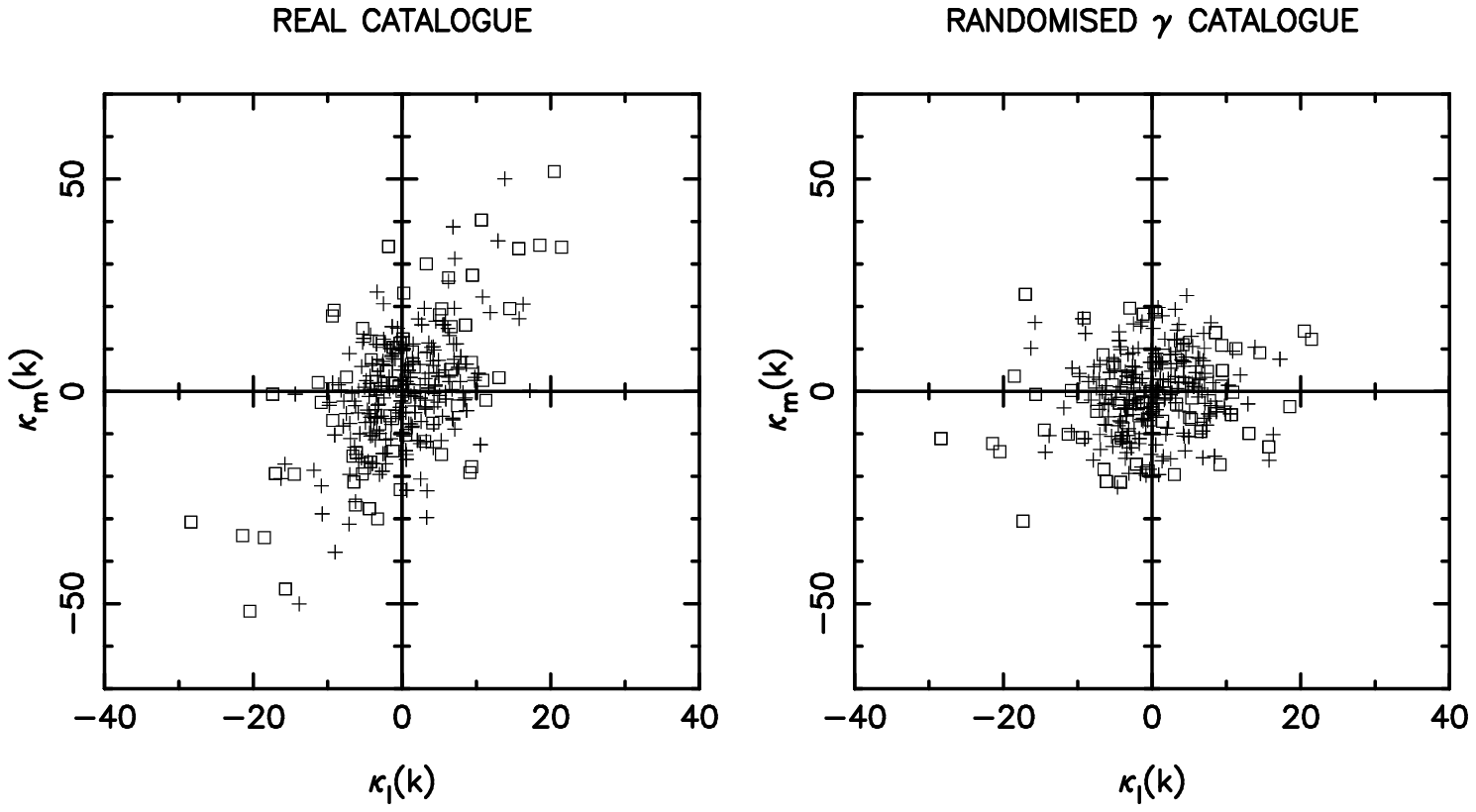,width=\figwidth,angle=-90}
\caption[Mass-light cross-correlation in Fourier Space.]{
Cross-correlation of light with mass reconstruction
in Fourier space. 
Left panel shows that actual cross
correlation and right panel shows the result of cross correlating the light
with a $\kappa$ reconstruction generated from a catalog with
randomized (shuffled) shear values.
Box symbols are for modes with
$\lambda = 1.5-6 h^{-1}$Mpc and $+$ symbols are for modes with wavelength 
$\lambda = 0.75-1.5 h^{-1}$Mpc.}
\label{fig:fft}
\end{figure}

Another way to see if $M/L$ changes with scale is to
perform the correlation in Fourier space.
The discrete Fourier transform of the $\kappa$ image provides a set of
estimates of the actual mass transform which have essentially statistically
independent measurement errors. 
The errors are exactly independent only if the data fully cover the
field - here there are some gaps in coverage like the N-W corner
which introduce slight correlations between neighboring
Fourier coefficients.  This slight covariance is properly accounted for 
by our Monte-Carlo error estimation. 
The results are shown in figure \ref{fig:fft}. Here the transform was taken
without zero padding.
If we use all modes with $k < 8$ times the fundamental frequency
for our square box  (corresponding to wavelengths $\lambda = 0.75-6 h^{-1}$Mpc)
we obtain a correlation which is significant
at the the $8.3$-sigma level and we obtain 
$M/hL = 233 \pm 28$.  If we split the modes into low and high
frequency subsets we find 
$M/hL = 280\pm42$ for $0 < k < 4$ and
$M/hL = 176\pm38$ for $4 \le k < 8$ so
again, we find a slight, though barely significant, increase in
$M/L$ with scale.

%% file: clusterml.tex
\begin{table*}[htbp!]
\begin{center}
\begin{tabular}{cccccc}
\hline\hline
cluster & $r_{\rm{ap}}$/arcmin & $\overline{\kappa}$ & $\nu$ & $h^{-1}M/L$ & $hM/10^{12}M_\odot$ \\
\hline
CL-S & $1.5$ & $0.0318\pm0.0096$ & $3.3$ & $241\pm72$ & $37.9\pm11.4$ \\
CL-E & $1.5$ & $0.0455\pm0.0164$ & $2.8$ & $182\pm65$ & $54.2\pm19.6$ \\
CL-N & $1.5$ & $0.0758\pm0.0100$ & $7.6$ & $342\pm44$ & $90.5\pm11.9$ \\
E+N+S & $1.5$ & $0.1586\pm0.0234$ & $6.8$ & $263\pm38$ & $189.1\pm27.9$ \\
CL-S & $3.0$ & $0.0131\pm0.0066$ & $2.0$ & $296\pm149$ & $62.4\pm31.5$ \\
CL-E & $3.0$ & $0.0206\pm0.0065$ & $3.1$ & $193\pm61$ & $98.1\pm31.2$ \\
CL-N & $3.0$ & $0.0434\pm0.0081$ & $5.3$ & $466\pm87$ & $206.9\pm38.7$ \\
E+N+S & $3.0$ & $0.0764\pm0.0140$ & $5.4$ & $314\pm57$ & $364.7\pm66.9$ \\
\hline
\end{tabular}
\caption[Cluster masses]{Cluster masses and mass to light ratios. The mean dimensionless surface density
$\overline\kappa$, the mass $M$, and the mass to light ratio have been corrected for
finite field effects.  The $\nu$ value is the significance of the measurement
determined by comparing with the rms mass measured for the ensemble of
randomised catalogs.
}
\label{tab:clusterml}
\end{center}
\end{table*}

%% file: discussion.tex
\section{Discussion}
\label{sec:discussion}

The foregoing analysis shows in a number of ways that we
have clearly detected the gravitational shear from 
mass concentrations in this field.  All of the major
concentrations apparent in the X-ray and optical images are
detected in the mass reconstructions, as well as a
foreground structure at $z \simeq 0.3$.
While the mass reconstructions are somewhat noisy, we
have shown by a number of different (though not independent)
methods that there is a strong correlation between the observed
shear and that predicted assuming early type galaxy light traces
the mass, which allow fairly precise estimate of the total
mass associated with these galaxies, and we find that
the $M/L$ parameter does not vary much with scale.

The dynamical state of the $z \simeq 0.42$ super-cluster is that most of the
mass in this region is concentrated into the three clusters.
Following \citeN{fbm94} we have modeled CL-S as a test particle on a
radial orbit around CL-N and CL-E as a test particle on a radial
orbit under the combined attraction of CL-N and CL-S.  We find generally
similar results, but as our masses our somewhat lower than they assumed
(by scaling to Coma using the velocity dispersion), our orbital
solutions are less tightly bound.  We find that CL-S is only marginally
bound to CL-N.  Equivalently, the mass of CL-N, if spread over a sphere
or radius equal to the distance to CL-S, gives a density about
equal to that of a critical density Universe at that epoch.  We find that
CL-E is on an unbound trajectory, though the latter conclusion could
possibly be modified if it turns out that CL-N, CL-S have
massive neighbors outside of the field we have surveyed.
While the clusters are bound and have collapsed, the super-cluster
as a whole is still rapidly expanding and may never turn around.

As remarked earlier, the conclusion 
that early type galaxies trace the mass is rather surprising.
The conventional picture of the morphological segregation
in superclusters is heavily influenced by the 
Local Supercluster.  There one finds \cite{bts87}
a E-rich core of dimension $\sim 1 h^{-1}$Mpc surrounded by a 
spiral rich `halo' extending to large radii ($\sim 10 h^{-1}$Mpc) 
roughly as $1/r^2$ and which dominates the total
light.  This picture is supported in a statistical sense
by the \citeN{dressler80} morphology-density relation and also
by galaxy correlation studies \cite{lmep95}.
It is commonly assumed either that galaxies in general trace
the mass, or, in biased theories and in hot
dark matter models, that the mass distribution
is even more extended.  What we have found is in conflict with either 
 of these pictures; the mass around the clusters in this field
is no more extended than the early type galaxies.

Exactly what this implies for the global density parameter $\Omega$
remains somewhat unclear because of the possibility of bias; as with other dynamical
estimates of cluster masses one can estimate $M/L$ and then extrapolate to
the entire Universe as a simple accounting exercise
to obtain the mass density as the product of $M/L$ with the
luminosity density measured from redshift surveys, but there is of course
no guarantee that the cluster $M/L$'s are really representative.
Nonetheless, and with the foregoing {\it caveat\/}, it is of interest to
make this extrapolation to obtain $\Omega$ 
assuming that galaxy formation is unbiased and
one can then say what level of bias is implied for
other values of $\Omega$.  

If we assume that galaxies of all types formed with their usual abundance
in this region then we have to ask what has happened to the late types.
One possibility is that they are still there and are more broadly
distributed than the early types as is the case in other `normal'
systems (we know from the spectroscopic studies that they
are not in the clusters themselves).  
Our data do not really support this picture, but we cannot firmly
rule it out either.  If so, these galaxies have very little
mass associated with them, and the total density of the Universe
is essentially just the luminosity density in early type galaxies times our $M/L$
and is very small.  \citeN{lpem92}, for example, find a total
luminosity density of $\simeq 2 \times 10^8h L_\odot / {\rm Mpc}^3$
of which about 20\% is due to the early types, and we thereby find 
$\Omega \simeq 280 \times 0.2 / 1500 \simeq 0.04$.  
The other possibility is that the late type galaxies in this region
have, like the early types, fallen into the clusters, but have
had their disks extinguished by ram pressure.  If there were much 
mass associated with these galaxies then we should have seen very
high mass-to-light ratios for these clusters, but we didn't, so
again this would suggest there is very little mass associated with
late types.
However, the bulges of the spirals will survive, and would have
inflated the cluster luminosity functions at the faint end
and the true value of $\Omega$ is then slightly higher.

If galaxy formation was unbiased then, our results suggest a very low
value for the density parameter, and, if $h$ is at the lower end
of the allowed range, this is mostly or all baryonic.
We do not know that early type galaxies are unbiased tracers, but the
assumption that they are is at least consistent with the general
idea that they formed early (so any statistical bias they may
have had a the time of formation may have been washed out) and that they
seem to be evolving slowly, and therefore seem to be obeying the
continuity equation in an average sense.
To avoid this rather radical conclusion one would have to invoke 
a bias.  However, it would seem to be very hard to accommodate these
data with say a flat EdS model, since in that case one has a region
with net mass density at
about or even slightly below the mean, but where the formation
of early type galaxies was apparently biased upwards by a factor 20 or so.
This seems unreasonable.  More modest positive bias cannot be ruled out, 
and it is also possible that the early type galaxies are `anti-biased';
i.e.~the abundance of these galaxies per unit mass is lower than average
in clusters, in which case the global $\Omega$ is still lower than the
above estimate.

We note that our mass determination is blind to any 
uniform density  component, so
we cannot rule out the possibility of a contribution to
$\Omega$ from very hot dark matter or
an ultra-light scalar field, the latter having
an effective Jeans length on the
order of the geometric mean of the horizon size and the Compton wavelength
and which could be very large. 

While the density parameter we find seems low as compared to other
estimates from e.g.~virial analysis (\citeNP{dp83};
\citeNP{cnoc96}), this is not because our
cluster $M/L$ values are much lower, it is because we do not extrapolate to
the Universe at large assigning the same $M/L$ to late type galaxies.
The interesting new conclusion from our results is that if late types 
do have the same net $M/L$ as
early types --- so $\Omega \simeq 0.2-0.3$ --- 
then their formation in the rather large region of
space originally occupied by the matter now in the supercluster 
must have been strongly suppressed
and the galaxy formation process must have arranged to create essentially
only early type galaxies, and with an abundance several times higher than
average to compensate.

Our cluster $M/L$s are considerably lower than that 
obtained from weak lensing
analyses of super-massive clusters like A1689 (e.g.~\citeNP{ft97}), 
and it is interesting to note that
the most massive of the three clusters here has a higher $M/L$ than 
the other two. 
If these results
are indications of a general trend then this would seem to argue against a
positive bias for clusters, since the most extreme clusters seem to
be less positively biased than more common poorer clusters like those
studied here. Equally if $\Omega$ is really $0.2-0.3$ as often assumed
then this requires non-monotonicity of bias with both
extremely massive clusters and field galaxies being essentially unbiased,
but with intermediate mass clusters being positively biased. This seems
contrived.
Finally, it is also interesting to briefly compare our results with
peculiar velocity analyses which provide an alternative probe of the
total (dark plus light) mass distribution.  Barring some very hot or
ultra-light mass component, our low $\Omega$ would seem to be very
hard to reconcile with the generally high values found from
large-scale `bulk-flows' \cite{dekel94}.  On the other hand, if there is
really very little mass outside of clusters, as our data suggest, then this
would be compatible with the very cold nature of the Hubble flow in
the field (\citeNP{sandage86}; \citeNP{bp87}; \citeNP{os90}).

%% file: acknowledge.tex
\section{Acknowledgements}
\label{sec:acknowledgements}

We thank John Tonry, Pat Henry and Harald Ebeling for
many helpful conversations and practical aid.
GW gratefully acknowledges financial support from
the estate of Beatrice Watson Parrent, from Mr. \& Mrs. Frank W.
Hustace, Jr., and from Victoria Ward, Limited.
This work was supported by NASA grant NAG5-7038,
NSF grant AST95-00515, NASA grants NAG5-1880 and NAG5-2523,
and CNR-ASI grants ASI94-RS-10 and ARS-96-13.
H.D. acknowledges support from 
a research grant provided by the Research Council of Norway, 
project number 110792/431.